\documentclass[12pt, preprint]{aastex}%

\def\mincir{\raise -2.truept\hbox{\rlap{\hbox{$\sim$}}\raise5.truept \hbox{$<$}\ }}
\def\mincireq{\hbox{\raise0.5ex\hbox{$<\lower1.06ex\hbox{$\kern-1.07em{\sim}$}$}}}
\def\magcir{\raise-2.truept\hbox{\rlap{\hbox{$\sim$}}\raise5.truept \hbox{$>$}\ }}

\newcommand{\bl}{\begin{list}{}}

\newcommand{\el}{\end{list}}

\newcommand{\magic}{\mbox{MAGIC}}
\newcommand{\iact}{\mbox{{IACT}}}
\newcommand{\iacts}{\mbox{{IACTs}}}

\newcommand{\vhe}{\mbox{VHE}}
\newcommand{\gray}{\mbox{$\gamma$-ray}}
\newcommand{\grays}{\mbox{$\gamma$-rays}}
\newcommand{\xrays}{\mbox{X-rays}}

\newcommand{\etal}{\mbox{et al}}

\newcommand{\meter}{\mbox{m}}

\newcommand{\fadc}{\mbox{FADC}}

\newcommand{\degrees}{$^{o}$}

\newcommand{\asm}{\mbox{{\it RXTE}/ASM}}
\newcommand{\nsb}{\mbox{NSB}}
\newcommand{\ZA}{\mbox{ZA}}

\newcommand{\eV}{\mbox{eV}}
\newcommand{\keV}{\mbox{keV}}

\newcommand{\TeV}{\mbox{TeV}}

\newcommand{\ns}{\mbox{ns}}

\newcommand{\cu}{\mbox{c.u.}}

\newcommand{\xray}{\mbox{X-ray}}
\newcommand{\Fvar}{\mbox{$F_{var}$}}

\newcommand{\gapp}{\ensuremath{\stackrel{\scriptstyle >}{{}_{\sim}}}}

\begin{document}
\title{%
Variable \vhe\ $\gamma$-ray emission from Markarian 501 \\ 
}%

%
\author{
 J.~Albert\altaffilmark{a}, 
 E.~Aliu\altaffilmark{b}, 
 H.~Anderhub\altaffilmark{c}, 
 P.~Antoranz\altaffilmark{d}, 
 A.~Armada\altaffilmark{b}, 
 C.~Baixeras\altaffilmark{e}, 
 J.~A.~Barrio\altaffilmark{d},
 H.~Bartko\altaffilmark{f}, 
 D.~Bastieri\altaffilmark{g}, 
 J.~K.~Becker\altaffilmark{h}
% ...
}
 \altaffiltext{a} {Universit\"at W\"urzburg, D-97074 W\"urzburg, Germany}
 \altaffiltext{b} {IFAE, Edifici Cn., E-08193 Bellaterra (Barcelona), Spain}
 \altaffiltext{c} {ETH Zurich, CH-8093 Switzerland}
 \altaffiltext{d} {Universidad Complutense, E-28040 Madrid, Spain}
 \altaffiltext{e} {Universitat Aut\`onoma de Barcelona, E-08193 Bellaterra, Spain}
 \altaffiltext{f} {Max-Planck-Institut f\"ur Physik, D-80805 M\"unchen, Germany}
 \altaffiltext{g} {Universit\`a di Padova and INFN, I-35131 Padova, Italy}  
 \altaffiltext{h} {Universit\"at Dortmund, D-44227 Dortmund, Germany}
% ...


\begin{abstract}
The blazar Markarian 501 (Mrk 501) was observed at energies above 0.10 TeV with the MAGIC telescope from May through July 2005. The high sensitivity of the instrument enabled the determination of the flux and spectrum of the source on a night-by-night basis. Throughout our observational campaign, the flux from Mrk 501 was found to vary by an order of magnitude. Intra-night flux variability with flux-doubling times down to 2 minutes was observed during the two most active nights, namely June 30 and July 9. These are the fastest flux variations ever observed in Mrk 501. The $\sim$20-minute long flare of July 9 showed an indication of a 4$\pm$1 min time delay between the peaks of F($<$0.25 TeV) and F($>$1.2 TeV), which may indicate a progressive acceleration of electrons in the emitting plasma blob. The flux variability was quantified for several energy ranges, and found to increase with the energy of the $\gamma$-ray photons. The spectra hardened significantly with increasing flux, and during the two most active nights, a spectral peak was clearly detected at 0.43 $\pm$ 0.06 TeV and 0.25 $\pm$ 0.07 TeV, respectively for June 30 and July 9. There is no evidence of such spectral feature for the other nights at energies down to 0.10 TeV, thus suggesting that the spectral peak is correlated with the source luminosity. These observed characteristics could be accommodated in a Synchrotron-Self-Compton (SSC) framework in which the increase in $\gamma$-ray flux is produced by a freshly injected (high energy) electron population.

\end{abstract}

\keywords{Markarian 501, BL~Lac, AGN, \vhe\ $\gamma$-ray, imaging air Cherenkov telescope, \magic}

\section{Introduction}
\label{Intro}

The large inferred luminosities of Active Galactic Nuclei (AGNs) led to a standard model 
of beamed AGN emission, with  the ultimate energy source being the release of gravitational 
potential energy of matter from an accretion disk surrounding a super-massive black hole 
\citep{Rees1984}. Particularly interesting for the Very-High-Energy \gray\ (\vhe \footnote{
  In this paper the \vhe\ band is defined 
  as the energy range $E$$\geq$0.1 TeV.})
community are the blazars, whose relativistic plasma jets point at the observer. 
Distinctive features of blazars are their continuum emission, clearly non-thermal from 
radio to \vhe\ frequencies and characterized by two broad bumps peaking at, respectively, 
IR/X-ray and \gray\ frequencies \citep{Blandford1978,UrryPadovani1995,ulrich1997}, and their 
strong variability, implying flux variations by a factor of $\magcir$10 over timescales of 
$\mincir$1 hour to months (flares; see Ulrich et al. 1997).

So far, 13 AGNs have been detected at \vhe\ energies. Except for M87 \citep{HegraM87, HessM87}, 
all these sources belong to the 'high-peaked BL Lac' (HBL) sub-class of blazars, which are
characterized by a spectral energy density (SED) in which both maxima occur at 
relatively high frequency (e.g., respectively at hard X-rays and HE/\vhe\ $\gamma$-rays).
The detection of \grays\ from blazars leads to some important considerations about the relevant 
radiation processes and the physical properties of the emitting regions. Most notably, the very 
detection of VHE radiation, implying \gray\ transparency in the emitting region, requires the 
presence of relativistic beaming to decrease the intrinsic energy density of the soft target 
photons inside the source. The beaming reduces this value because it simultaneously decreases the 
intrinsic energy density of the photons and, having reduced the energy of the relevant \gray\ 
photons, it increases the energy of the soft target photons relevant to e$^{\pm}$ production, 
hence (for typical spectra) decreasing their number density (McBreen 1979; \citep{Blandford1979}.
Bassani \& Dean 1981; \citep{Mattox1993}; Dondi \& Ghisellini 1995). 

Two classes of emission models have been proposed to explain the \TeV\ emission from blazars: 
leptonic and hadronic models. \\
\noindent
{\it (i)} In the case of the most popular leptonic models, the same population of non-thermal 
electrons (and possibly positrons) responsible for the radio-to-\xrays\ SED is also responsible 
for \gray\  emission, through Compton up-scattering of the synchrotron photons off their own 
parent electrons -- the Synchrotron-Self-Compton (SSC) process \citep{marscher1985, maraschi1992, 
bottcher2002}. In other models, electrons scatter 'external' photons that originate outside the jet 
(External Compton (EC) models\footnote{
  The external seed photons may, e.g., come from 
  the accretion disk \citep{DermerSchlickeiser1993}, or be the 
  disk radiation scattered by the material around the disk and 
  and the jet \citep{Sikora1994}, or be radiation from the massive 
  stars which enter the jet \citep{Bednarek1997}, or be synchrotron 
  radiation produced in the jet and reflected by the surrounding 
  material \citep{Ghisellini1996}.}).
In BL Lacs the lack of strong emission lines suggests a minor role 
of ambient photons and hence supports the SSC models.\\
\noindent
{\it (ii)} In hadronic models, the \TeV\ radiation is produced by hadronic interactions of 
the highly relativistic baryonic outflow with the ambient medium\footnote{
I.e.: gas and clouds drifting across the jet \citep{laor1997,bednarek1999}, 
the matter from the thick accretion disk \citep{Bednarek1993}, 
or interactions inside the (dense) jet \citep{pohl2000}.},
and/or by interactions of ultra-high-energy protons with 
synchrotron photons produced by electrons \citep{manheim1992}, with the jet magnetic field 
\citep{aharonian2000}, or with both \citep{mucke2003,Atoyan2003,manheim1993}. \\
\noindent
However, hadronic models are challenged by the blazars' observed \xray\ versus \vhe\ correlation and 
very rapid \gray\ variability. The SSC model is then widely believed to explain the dominant emission 
process in blazar jets (not always in its simplest one-zone realization, as required by e.g. 'orphan 
flares': see \citet{Krawczynski2004} and \citet{gliozzi2006}). 

In this framework, the importance of high-quality data on blazar \vhe\ emission can not be overestimated. 
In particular, valuable information can be obtained investigating: 
{\it (i)} the rapid, possibly energy-dependent, flux variability;
{\it (ii)} the \xray /optical versus \vhe\ correlation; and 
{\it (iii)} the \xray\ and \vhe\ spectral variability, 
with a potential energy shift of the Synchrotron and Compton peaks.
Simultaneous multi-wavelength observations 
of such rapid variability can provide stringent tests to emission models, in particular on 
acceleration processes in jets. In the case of nearby 
($z$$<$0.1) sources, the extinction due to pair production by interaction of the 
blazar-emitted \TeV\ photons with the (not well known) optical/IR Extragalactic Background Light (EBL) photons 
is probably minor, and thus there is a smaller uncertainty in the determination of the 
intrinsic spectral features of the object.

In this paper we report about detailed measurements of the \vhe\ emission of \mbox{Mrk 501},
demonstrating the capability of the \magic\ Telescope in the precision study of 
blazar physics.
The BL Lac object \mbox{Mrk 501} was the second established \TeV-blazar \citep{Quinn1996, 
Bradbury1997}. After a phase of moderate emission for about a year following its 
discovery as a TeV source, in 1997 \mbox{Mrk 501} went into a state of surprisingly high 
activity and strong variability, becoming $>$10 times brighter (at energies $>$1 
TeV) than the Crab Nebula \citep{AharonianMrk501-1999a, AharonianMrk501-1999b}. 
In 1998-1999 the mean flux dropped by an order of magnitude \citep{AharonianMrk501-2001}. 
It is worth noticing that the HEGRA observations (with a threshold energy of $\sim$0.5 \TeV) 
did not see spectral variations during the 1997 outburst, whereas it did 
observe in 1998-1999 a significantly softer low-state energy spectrum than in 1997. 
The CAT telescope (with a threshold energy of $~$0.25 \TeV), on the other hand, did 
detect spectral variations during 1997 \citep{Mrk501CAT,Mrk501CATPiron,Mrk501CATPironMoriond}.

The structure of the paper is as follows. In sect. 2 we briefly describe the 
\magic\ data taking and analysis. In sects. 3 and 4 we present and
discuss the Mrk 501 \vhe\ 
lightcurve (hereafter LC), spectrum and their variability, during the observation campaign. 
In sect.5 we discuss the long-term light curve of Mrk 501, 
short-term flux variability, flux-spectrum 
correlations, and the overall SED of this object.
Finally, sect. 6 summarizes our main results.

\section{The \magic\ Telescope and the data analysis}
\label{Obsdata}

\subsection{The instruments}
\label{Magic}

The observations in the \vhe\ domain were carried out with the Major Atmospheric Gamma-ray Imaging 
Cherenkov (\magic) telescope, located on the Canary island of La Palma (28.8\degrees $N$, 17.9\degrees 
$W$) at the \textit{Roque de los Muchachos Observatory} (about 2200 \meter\ above see level). 
\magic\ started regular observations in the fall of 2004 and, with a main mirror diameter of \mbox{17 \meter}, 
it is currently the world's largest single-dish \iact.
Further details about the characteristics and performance of \magic\ can be 
found elsewhere \citep{Baixeras2004,PanequeThesis,Cortina2005,GaugThesis}.

The \magic\ Collaboration also operates the optical KVA telescope \mbox{(35 cm)}. 
Simultaneously with the 
\magic\ observations, Mrk 501 was regularly observed with KVA as a part of 
the Tuorla Observatory blazar monitoring program\footnote{
\tt http://users.utu.fi/kani/1m/.}. 
In this paper we also use 2-10 \keV\ data taken with the 
{\it RXTE} satellite's All-Sky-Monitor (\asm)\footnote{
The data are publicly available at \mbox{\tt http://heasarc.gsfc.nasa.gov/xte\_weather/}.}
quasi-simultaneously with our \magic observations.

\subsection{Source observation}
\label{Observation}

The source was observed during 30 nights between May and July 2005, with an overall observation time 
of 54.8 hours. In order to maximize the time coverage of this source, observations were carried out 
also in the presence of moonlight (34.1 hours, i.e. 62\% of the total observing time). It is important 
to note that many of these 'moon observations' were performed when the moon was only partly illuminated, 
and mostly located at a large angle (60\degrees-90\degrees) with respect to the position of Mrk 501. This 
kept the Night Sky Background (\nsb) of these observations rather low and comparable to that of the 
moonless observations. The observations were mostly performed in the so-called 'on mode' in which the 
telescope points exactly to the source ({\it on-data}), and thus its (optical) image is right in the 
center of the camera. The telescope was also operated in 'off mode', in which it
points to regions of the sky where there are no known 
\gray\ sources ({\it off-data}). These observations were 
carried out at comparable zenith angle (\ZA) and \nsb\ conditions, and can therefore be used to estimate 
the background content of the {\it on-data}. The {\it off-data} observation time amounts to 3.5 hours.

The data were screened for hardware problems, non-optimal weather conditions, and too high \nsb\ light. 
In addition, the very few runs with \ZA\ larger than 30\degrees\ that survived those filters were also 
removed, in order to have a more uniform data set. The number of nights surviving these selection cuts is 
24, with a total net observation time of 31.6 hours\footnote{
Five out of the six entirely rejected observation 
nights (with a total observation time of 9.6 hours), 
were moonlight observations.}. 
These observation nights, together with the corresponding net times and \ZA\ ranges of the data 
acquisition, are listed in Table \ref{SummaryTable}.

\subsection{Data analysis}
\label{DatasAnalysis}

The analysis used in this paper is based on the Hillas image parameters 
ALPHA, WIDTH, LENGTH, DIST and SIZE to quantify the camera image, see \citet{HillasICRC1985}. 
These parameters are calculated using the calibrated signals from the individual pixels of 
the camera. The procedures used to calibrate the PMT signals in \magic\ are described in 
detail in \citet{GaugThesis} and \citet{MagicSigRecon2007}: here we used the {\it Sliding Window} 
signal extractor, and performed the calibration with the {\it Excess Noise Factor} method. 
Only PMT signals with more than 10 photoelectrons (8 photoelectrons for pixels in 
the boundary of an image) which occur within a time window of 6.6 \ns\ (2 \fadc\ slices) of 
the neighbouring pixel signal were used. The minimum total image light content (SIZE) considered in 
this analysis is 150 photelectrons. 
The {\it signal/background} separation is achieved by applying 
dynamical cuts (defined as 2nd order polynomial functions of logSIZE) on the parameters WIDTH, LENGTH 
(shape parameters) and DIST (position of the image).  The background 
contained in the {\it on-data} after the $\gamma$/hadron separation cuts is estimated by 
means of a 2nd-order polynomial fit (with no linear term) to the ALPHA distribution from the 
normalized (according to the on-off observation times) {\it off-data}. 

The energy of the incoming \grays\ is essentially proportional to the light content of the 
image (SIZE), with corrections according to the values of LENGTH, DIST and LEAKAGE\footnote{
LEAKAGE is defined as the fraction of the light 
content recorded by the outer ring of the PMT 
camera, and it is typically used to evaluate the 
level of missing light in the detected image.}. 
The energy resolution achieved with this parameterization is about 20-30\%, slightly 
depending on the event's energy. Because of the finite experimental resolution, the 
distribution of the excess events ($\gamma$-candidates) versus the reconstructed energy is a 
convolution of the (true) energy distribution of excess events and a realistic 
energy resolution function. The determination of the true energy spectrum from the 
reconstructed one is achieved by means of an unfolding procedure \citep{Anykeyev1991}. 
We used the iterative method described in \citet{Bertero1989}.

In this analysis, 90-99\% of the background images are removed by the selection cuts, while 
50-60\% of the \gray\ signals are kept.
The resulting collection area after analysis cuts is \mbox{$\gapp 0.5 \cdot 10^5 ~m^2$} 
down to 0.20 \TeV. 
The analysis threshold energy, commonly defined as the peak of the 
differential event-rate spectrum after all cuts, is $\sim$ 0.15 \TeV. 
The lowest \gray\ energy used in the calculation of the energy spectra is 0.10 \TeV.
In the LC analysis, however, the minimum energy considered is 0.15 \TeV\  (i.e. the threshold 
energy). Below this energy the collection area drops fast, increasing rapidly both 
systematic and statistical errors in the measured flux. Keeping the measurement errors small 
is essential for the study of flux variations, one of the main goals of this work.

In order to check the reliability of the used analysis chain, we analyzed data from the 
Crab Nebula taken in December 2005, under similar instrumental and environmental 
conditions to those of Mrk 501. The obtained results were in perfect agreement with 
data published previously \citep{CrabWhipple1998,CrabHegra2004,CrabRobert2005}, 
which shows that the analysis procedures used produce reliable results. 

The results from the MAGIC \mbox{Mrk 501} data analysis are summarized in \mbox{Table 
\ref{SummaryTable}}. The table shows the integrated flux (above 0.15 \TeV) and the resulting 
fit to the differential (energy) photon spectra  with a simple power-law (PL) model (see 
eq. \ref{eq_powerlaw}), for each observing night. The fit was obtained using all the 
spectral points above 0.10 \TeV. Only statistical errors are reported in \mbox{Table 
\ref{SummaryTable}}. The systematic errors on the energy determination are estimated as 
$\sim$20\% which, for a spectral index of 2.5, would produce a systematic shift of 50\% in 
the flux level (normalization factor of the PL function from Table \ref{SummaryTable}). The 
systematic error in the calculated spectral indices is evaluated as $\sim$0.1. In the table 
we quote the combined significance which is calculated following the prescription given by 
\citet{combsign2006} as 
\begin{equation}
\label{eq_combsign}
S_{comb} = \frac{\sum S_i}{\sqrt{n}}
\end{equation}
where $S_i$ is the significance corresponding to the (differential) energy bin $i$, and $n$ is 
the number of energy bins (measurements). The significance of each energy bin is calculated 
according to eq.(5) of \citet{LiandMa}, which is more suitable than eq.(17) from the same 
paper, since in all nights the \gray\ signal is clear and hence its existence is not in doubt. 
The combined significance is used to compare the quality of the \gray\ signals from 
different observing nights.

\section{Lightcurve of \mbox{Mrk 501} during the \magic\ observations}
\label{Lightcurve}

In this section we report on the broadband, optical to \gray, LC of 
\mbox{Mrk 501} during May-July 2005. 

\subsection{Overall lightcurve at \gray, \xray, and optical frequencies}
\label{OverallLigthCurve}

\begin{figure}
\epsscale{1}
\plotone{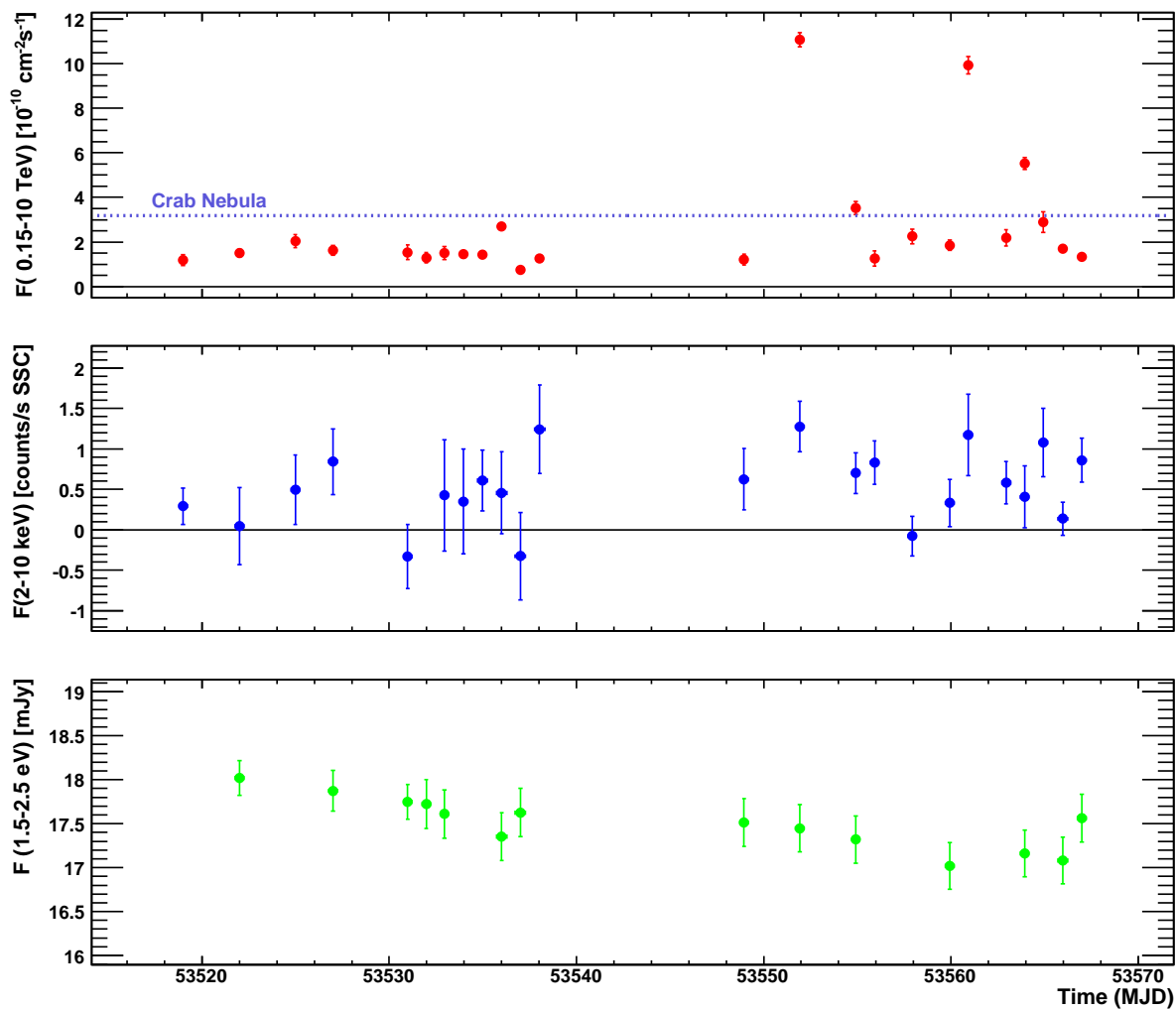}
\caption{Multi-frequency LC during the \magic\ observations of \mbox{Mrk 501} (May-July 2005). 
{\it Top)} \magic\ flux above 0.15 TeV. The Crab flux is also shown for comparison (lilac dotted
horizontal line). 
{\it Middle)} $RXTE$/ASM 2-10 keV flux.  
{\it Bottom)} KVA $\sim$1.5-2.5 eV flux. 
Error bars denote 1$\sigma$ statistical uncertainties. The X-ray/optical data were selected to match 
the MAGIC data within a time window of 0.2 days.
\label{OverallLightCurve}}
\end{figure}

The overall LC of \mbox{Mrk 501} during the \magic\ observation campaign is shown in Fig. 
\ref{OverallLightCurve}. The observed flux is shown in three energy bands: \vhe\ 
(0.15\TeV-10\TeV), \xrays\ (2\keV-10\keV), and optical (1.5\eV-2.5\eV) as 
measured by \magic, \asm\, and KVA, respectively. The \xray\ and optical fluxes are computed 
as weighted averages using \asm\ and KVA measurements taken simultaneously with the \magic\ 
observations plus/minus a time tolerance  of 0.2 days. A smaller time tolerance substantially 
decreases the number of \xray\ points that can be used. The flux level of the Crab Nebula 
(lilac-dashed horizontal line in the top plot) is also shown in Fig. \ref{OverallLightCurve} 
for comparison. The Crab Nebula flux was obtained by applying the very same analysis as 
described in Sect. \ref{DatasAnalysis} to the MAGIC Crab Nebula data taken during December 
2005 under observing conditions similar to those for Mrk 501. The estimated Crab Nebula flux 
level is therefore roughly affected by the same systematics as the fluxes obtained for Mrk 501. 
We found $F_{\rm Crab}$($>$$0.15\,$TeV)=(3.2$\pm$0.1)$\times$$10^{-10}$ cm$^{-2}$s$^{-1}$, 
thereafter referred to as Crab Unit (\cu). For simplicity, only the Crab Nebula flux level, 
and not the associated error (which is irrelevant for the comparison) is shown in the LC.

The measured \vhe\ flux from \mbox{Mrk 501} was at about 0.5 \cu\ during most of the observation 
nights (Table \ref{SummaryTable}). During several nights, however, its flux significantly exceeded 
0.5 \cu, and during one night (MJD 53536.947) it showed a substantially lower flux (\mbox{0.24$
\pm$0.04 \cu}). Often, Mrk501 showed large flux variations in consecutive nights. 
An example of these rapid flux variations are the MJD 
53535.934 and 53536.947 with respective fluxes of \mbox{0.84$\pm$0.04 \cu} and \mbox{0.24$\pm$0.04 
\cu}; the MJD 53554.906 and 53555.914 with respective fluxes of \mbox{1.11$\pm$0.09 \cu} and 
\mbox{0.40$\pm$0.11 \cu}; and the MJD 53563.921 and 53564.917 with respective fluxes of \mbox{1.74$
\pm$0.09 \cu} and \mbox{0.91$\pm$0.15 \cu}. Besides, the \vhe\ flux from \mbox{Mrk 501} was 
outstanding during the MJD 53551.905 (June 30) and 53560.906 (July 09) with \mbox{3.48$\pm$0.10 \cu} 
and  \mbox{3.12$\pm$0.12 \cu}, respectively. During these two nights the source was in a very active 
state. Note, however, that the night before the July 9 flare the emitted flux was 0.58$\pm$0.07 \cu, 
i.e. close to the average flux of the entire campaign. \mbox{Mrk 501} therefore showed a remarkably 
fast VHE variability during this campaign.

Unlike in VHE \grays, no significant flux variation was recorded in the \xray\ and optical bands. In the 
case of the \xray\ data, however, although the sensitivity of the \asm instrument was clearly inadequate 
to reveal short-term 2-10 \keV\ flux variability in Mrk 501's emission, the flux appears to be higher in 
the second portion of the LC. The optical flux, on the other hand, shows only a modest variation, a 
$\sim$$5\%$ monotonic decrease during the entire observational campaign.

\subsection{Multi-frequency correlations}
\label{Correlation}

\begin{figure}
\epsscale{1}
\plotone{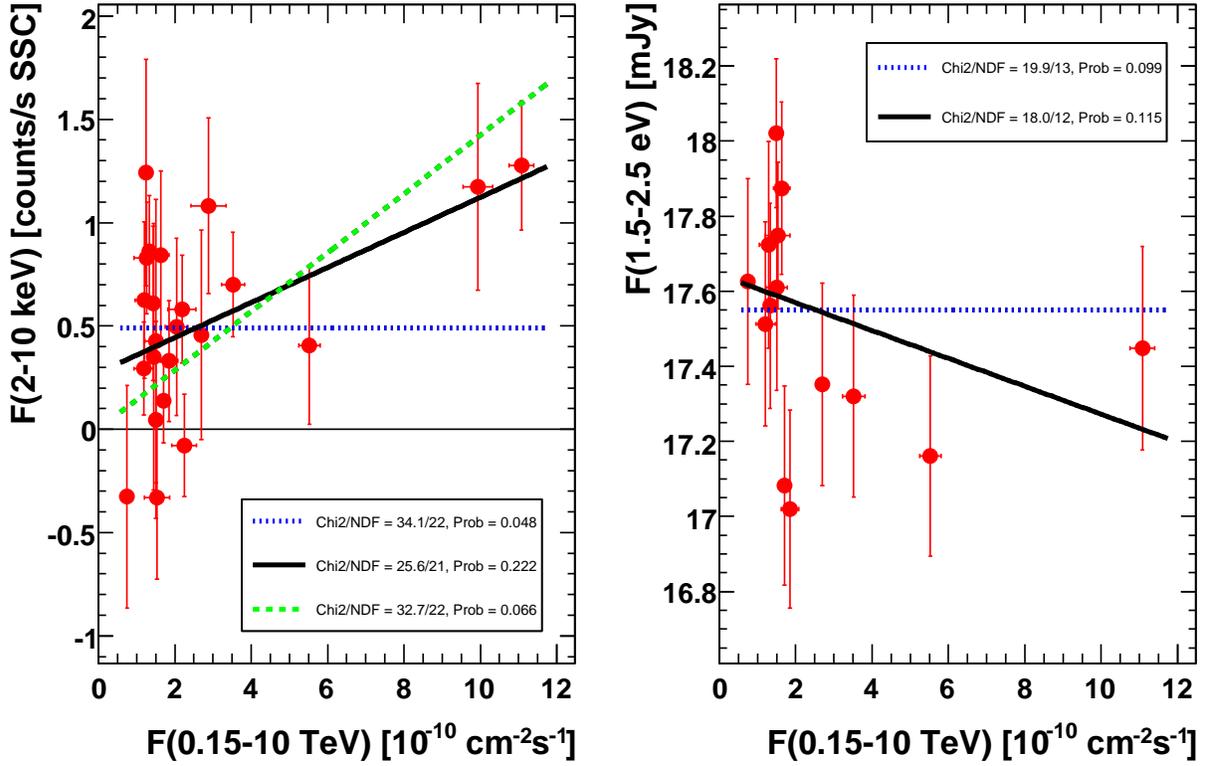}
\caption{\vhe\ versus X-ray ({\it left}) and optical ({\it right}) flux correlation during the 
\magic\ observation campaign. The data points are the same as in Fig. \ref{OverallLightCurve}. The blue 
dotted lines denote constant fits; black solid and green dashed lines correspond to linear fits with/without offset (see 
insets for goodness-of-fits values).
\label{CorrGammaXrayOptical}}
\end{figure}

The correlations of our observed VHE \gray\ data with X-ray and optical data are 
shown in Fig. \ref{CorrGammaXrayOptical}, the gamma points being the same as shown in the 
light curve of Fig. \ref{OverallLightCurve}. It can be seen that the measurement 
uncertainties of the X-ray and optical fluxes are comparatively large, 
which makes possible different dependencies of the X-ray flux on the  \gray\ flux. 
We fit the VHE/X-ray data with a constant and with a linear function, obtaining the 
highest probability for the linear function.
For the relation VHE/optical data, the two fits are nearly equally probable. 
All fit results are shown in the insets of Fig. \ref{CorrGammaXrayOptical}, 
including for the VHE/X-ray data also 
a fit forcing the linear function through the origin.

\begin {figure} 
\epsscale{0.5}
\plotone{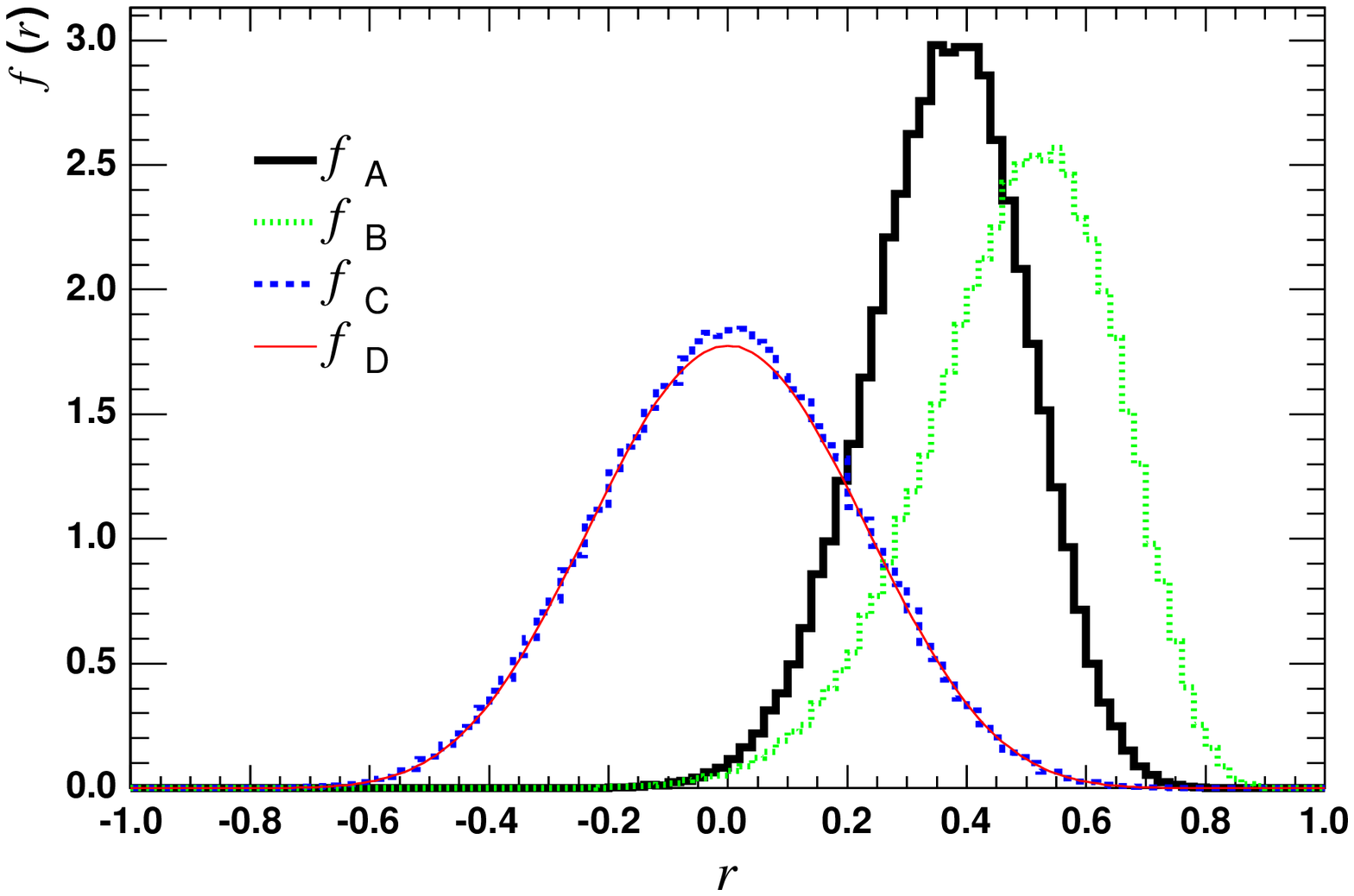}
\caption{Probability density
functions of the correlation coefficient between \gray\ and \xray\ fluxes: $f_A$ data, $f_B$ perfectly 
correlated case, $f_C$ uncorrelated case, and $f_D$ analytical solution in the uncorrelated case. See 
text for further details.
\label {GammaXCorr}}
\end {figure}

For the VHE/X-ray correlation, one obtains a linear correlation coefficient of 0.49.  
Investigating the uncertainty of this result, we used a procedure of Monte Carlo-generated 
correlations, as described in detail by \citet{Fer07,Hru07}. 
For all points (corresponding VHE and X-ray fluxes), multiple (we used 100 000) 
possible sets of measurements are generated, using random differences derived from the (Gaussian) 
errors assigned to each point. For each set of generated points, 
a correlation coefficient is obtained, resulting for a large number of measurements in 
the probability density function (pdf) of correlation coefficients $f_A(r)$ shown in 
Fig. \ref{GammaXCorr}, which corresponds to the measured correlation and the assigned errors. 
This pdf correctly expresses the effect of the measurement uncertainties.
The same procedure can be applied to hypothetical fully 
correlated data: to this end, the data points are shifted onto the straight line 
from the fit with the highest probability 
(black line in left-hand plot of Fig. \ref{GammaXCorr}), maintaining the 
original error assignments. In the case of hypothetical uncorrelated data, the 
points are randomly distributed in the  VHE/X-ray plane.
Sets of Monte Carlo measurements are then generated as before. 
The resulting pdfs express the probability to obtain certain values of correlation coefficients, 
given respectively no correlation ($f_C(r)$ in Fig. \ref{GammaXCorr}) 
or full correlation ($f_B(r)$ in Fig. \ref{GammaXCorr}) with our error assumptions. 
For comparison purposes, Fig. \ref{GammaXCorr} also shows the analytical pdf function 
for the uncorrelated case, $f_D(r)$ (described in chapter 9 and appendix C of \citep{Tay97}), 
which does not take into account measurement errors.  
Note that $f_C(r)$ is very similar to $f_D(r)$. This is not 
surprising, since the smearing does not increase the randomness of the already randomized seed event. 
The width of the probability density distribution $f_D(r)$ depends exclusively on the number of points 
in the data sample (23 in our case). 
On the other hand, $f_B(r)$ is significantly affected by the measurement errors; 
even under the hypothesis of a fully correlated case, 
the probability to obtain values for the correlation coefficient larger than 0.8 is very small.

A measure of the probability of correlation can be derived from the comparison of the probability 
density distribution $f_A(r)$ for the actual measurement with the distributions for the two 
extreme correlation cases, $f_B(r)$ and $f_C(r)$. Given the results in Fig. \ref{GammaXCorr}, 
it is evident that $f_A(r)$ is similar to $f_B(r)$, but, despite a sizable overlap, rather 
different from $f_C(r)$. For a quantitative comparison, we followed the robust method from 
\citet{Poe05}, which is based on the convolution of empirical probability density distributions. 
The comparison of a pair of probability density functions $f_X(r)$ and $f_Y(r)$ leads to the 
probability that these two distributions are statistically consistent, $P(f_X(r),f_Y(r))$. The 
resulting value for the probability of agreement between our data points and the 
fully correlated case, $P_{X,\gamma}(f_A,f_B)$, is $0.55 \pm 0.15$, being the quoted error a 
systematical error estimated through variations in the initial values used in the Monte Carlo 
method \citep{Fer07,Hru07}.  The 
probability that our measurement came as a result of a statistical fluctuation of an entirely 
uncorrelated physics case, on the other hand, is significantly lower; $P_{X,\gamma}(f_A,f_C)$ 
is $0.15 \pm 0.05$. The probability for the first scenario to be true {\it and} for the second 
to be false is $P_{X,\gamma}(f_A,f_B)(1 - P_{X,\gamma}(f_A,f_C)) = 0.47$, and the probability
for the opposite is $(1 - P_{X,\gamma}(f_A,f_B)) P_{X,\gamma}(f_A,f_C) = 0.067$, which indicates 
that the correlation scenario is significantly more likely\footnote{
It is worth noticing that the present analysis 
treats data from low and very-high activity epochs 
together, whereas the same correlation slope may not 
be necessarily the same for the two activity states.}.

\begin {figure} 
\epsscale{0.5}
\plotone{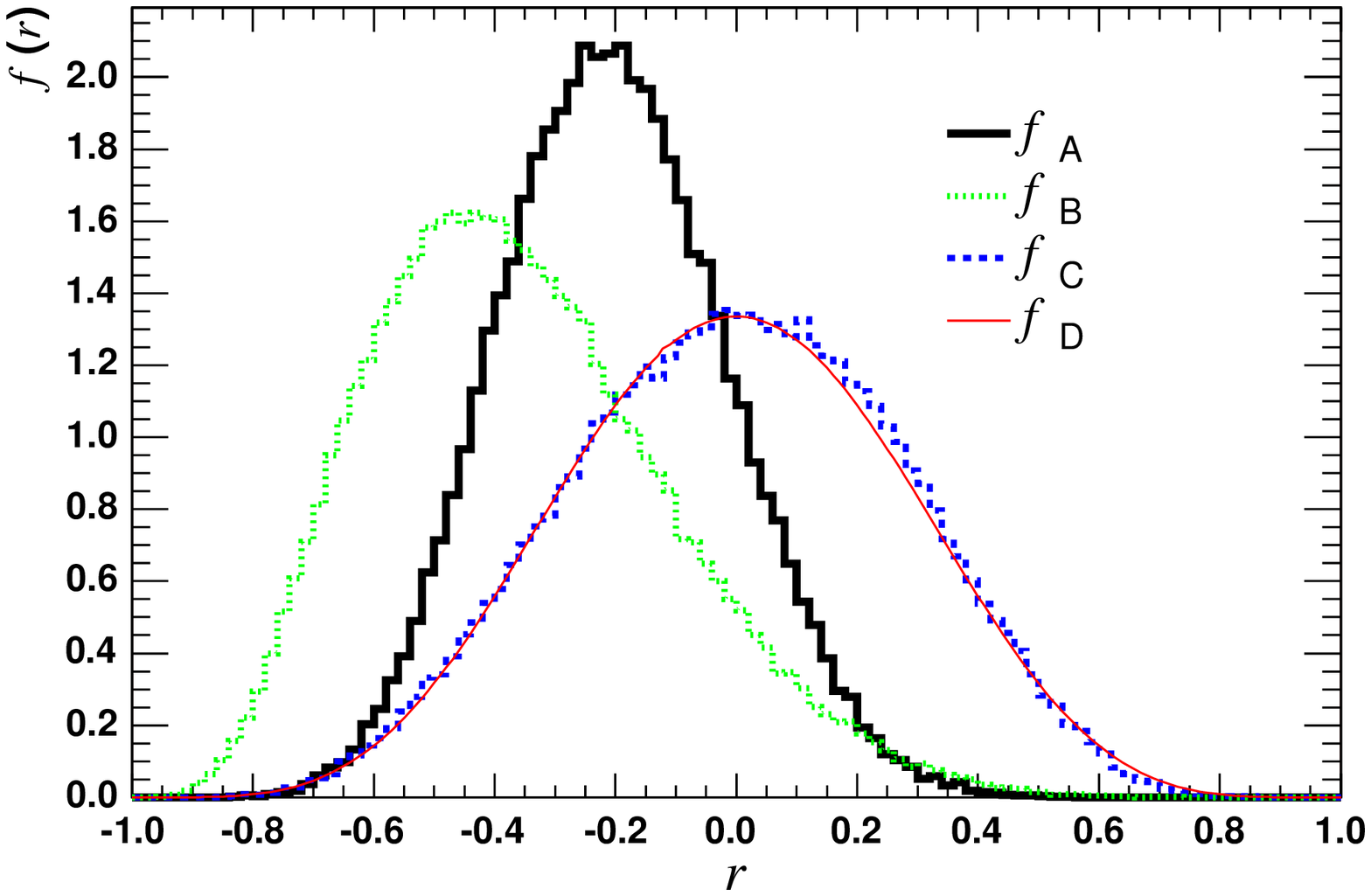}
\caption{Probability density
functions of the correlation coefficient between \gray\ and optical fluxes: 
$f_A$ data, $f_B$ perfectly correlated case, $f_C$ uncorrelated case, 
and $f_D$ analytical solution  in the uncorrelated case.
See text for further details.
\label {GammaOptCorr}}
\end {figure}

The same method was applied to the optical and \gray\ flux values of Fig. \ref{CorrGammaXrayOptical} 
(right-hand plot). The corresponding probability density functions are presented in Fig. \ref{GammaOptCorr}. 
In this case, the resulting linear correlation coefficient is -0.27, indicating a small anti-correlation;
yet the probabilities $P_{{\rm opt}, \gamma}(f_A,f_B)$$=$0.60$\pm$0.25, and $P_{{\rm opt},\gamma}(f_A,f_C)$=0.55$
\pm$0.25 are practically equal, which suggest that neither the anti-correlated, nor the uncorrelated scenario 
may be reliably excluded. The probability for the first scenario to be true {\it and} for the second to be 
false is 0.27, while the probability for the opposite case is 0.22, which confirms the previous conclusion.

\subsection{Intra-day \gray\ flux variations}
\label{Intradayvariations}

\begin{figure}
\epsscale{1}
\plotone{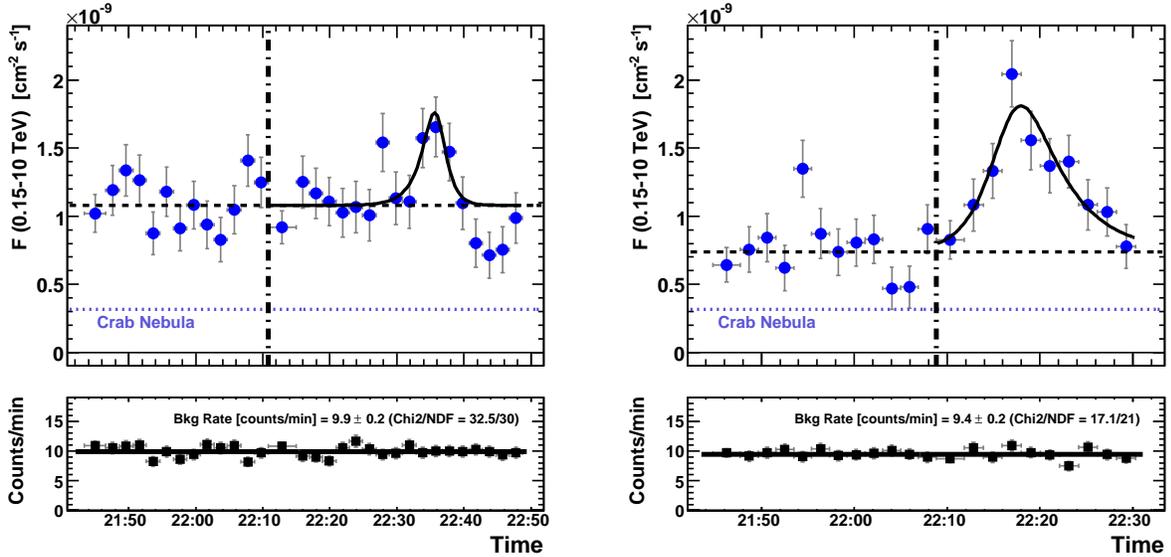}
\caption{Integrated-flux LCs of \mbox{Mrk 501} for the flare nights of June 30 and July 9. 
Horizontal bars represent the 2-minute time bins, and vertical bars denote 1$\sigma$ statistical 
uncertainties. For comparison, the Crab emission is also shown as a lilac dotted horizontal line. 
The vertical dot-dashed line divides the data into 'stable' (i.e., pre-burst) and 'variable' (i.e., 
in-burst) emission. The horizontal black dashed line represents the average of the 'stable' 
emission. The solid black curve represents the best-fit flare model (see eq. \ref{eq_fitflare}).
The bottom plots show the mean background rate during each of the 2-minute bins of the LCs.
The insets report the mean background rate during the entire night, resulting from a 
constant fit to the data points. The goodness of such fit is also given. 
\label{LCSingle}}
\end{figure}

During the two nights with the highest \vhe\ activity, namely June 30 and July 9, Mrk 501 
clearly showed intra-night flux variations. The corresponding LC in the 0.15-10 TeV band is 
shown in Fig. \ref{LCSingle} with a time binning of $\sim$2 minutes. For comparison, the Crab 
Nebula flux is shown as a lilac dashed horizontal line. The vertical dot-dashed line divides 
the data into a region of relatively 'stable' (pre-burst) emission and 
one of 'variable' (in-burst) 
emission. The 
background rate after the gamma/hadron selection cuts was evaluated during these two 
nights, and is shown in the bottom panels of Fig. \ref{LCSingle}. 
These rates were found to be constant along the entire night. Consequently, the variations 
seen in the upper panels of Fig. \ref{LCSingle} correspond to actual 
variations of the \vhe\ \gray\ flux from Mrk501, thus 
ruling out detector instabilities or atmospheric changes. 

A constant line fit to the whole LC gives a $\chi^2/NDF=47.9/30$ (probability $p=2.0 \times 
10^{-2}$) for the night of June 30, and  a $\chi^2/NDF=80.6/21$ ($p=6.4\cdot10^{-9}$) for the 
night of July 9. The emission above 0.15 \TeV\ during the two nights is therefore statistically 
inconsistent with being constant and it is more reasonable to only fit the first part, i.e. the 
'stable' emission, with a constant - and not the entire LC. A constant fit to the 'stable' 
portion of the LCs gives $\chi^2/NDF=13.4/12$ ($p$=0.34) for June 30, and $\chi^2/NDF=17.8/11$ 
($p$=0.09) for July 9 (see Fig. \ref{LCSingle}). The probability that the 'variable' parts of 
the LCs are compatible with the 'stable' flux level is given by $\chi^2/NDF=34.5/18$ ($p=1.1 
\times 10^{-2}$) for June 30 and $\chi^2/NDF=83.3/10$ ($p=1.1\cdot10^{-13}$) for July 9, 
respectively. We therefore measured intra-night flux variations in both nights.

The flare's amplitude and duration, as well as its rise/fall times, can be quantified 
according to 
\begin{equation}
\label{eq_fitflare}
F(t) = a + \frac{b}{2^{-\frac{t-t_0}{c}} + 2^{\frac{t-t_0}{d}}}
\end{equation}
\citep{thomasthesis}. 
This model parameterizes a flux variation (flare) superposed on a stable emission: $F(t)$ 
asymptotically tends to $a$ when $t \rightarrow \pm \infty$. The parameter $a$ is the assumed 
constant flux at the time of the flare (cf. the horizontal black dashed lines in Fig. 
\ref{LCSingle}); $t_0$ is set to the time corresponding to the point with the highest value in 
the LC; and $b,c,d$ are left free to vary. The latter two parameters denote the flux-doubling 
rise and fall times, respectively, and can be converted into the characteristic rise/fall 
times\footnote{
  The characteristic time is defined as the time 
  needed for the flux to change by $e^{\pm 1}$.} 
by multiplying them by $1/ln2$. The resulting fits using eq. \ref{eq_fitflare} are shown in Fig. 
\ref{LCSingle}, and their parameter values are reported in Table \ref{fitflareresults}. In both 
cases, the measured rise/fall flux-doubling times are $\sim$2 minutes, which yields 
characteristic rise/fall times of \mbox{2/ln2 $\sim$3 minutes.} These are the shortest 
flux-variation timescales ever measured from \mbox{Mrk 501}, at any wavelength.

\begin{figure}
\epsscale{0.7}
\plotone{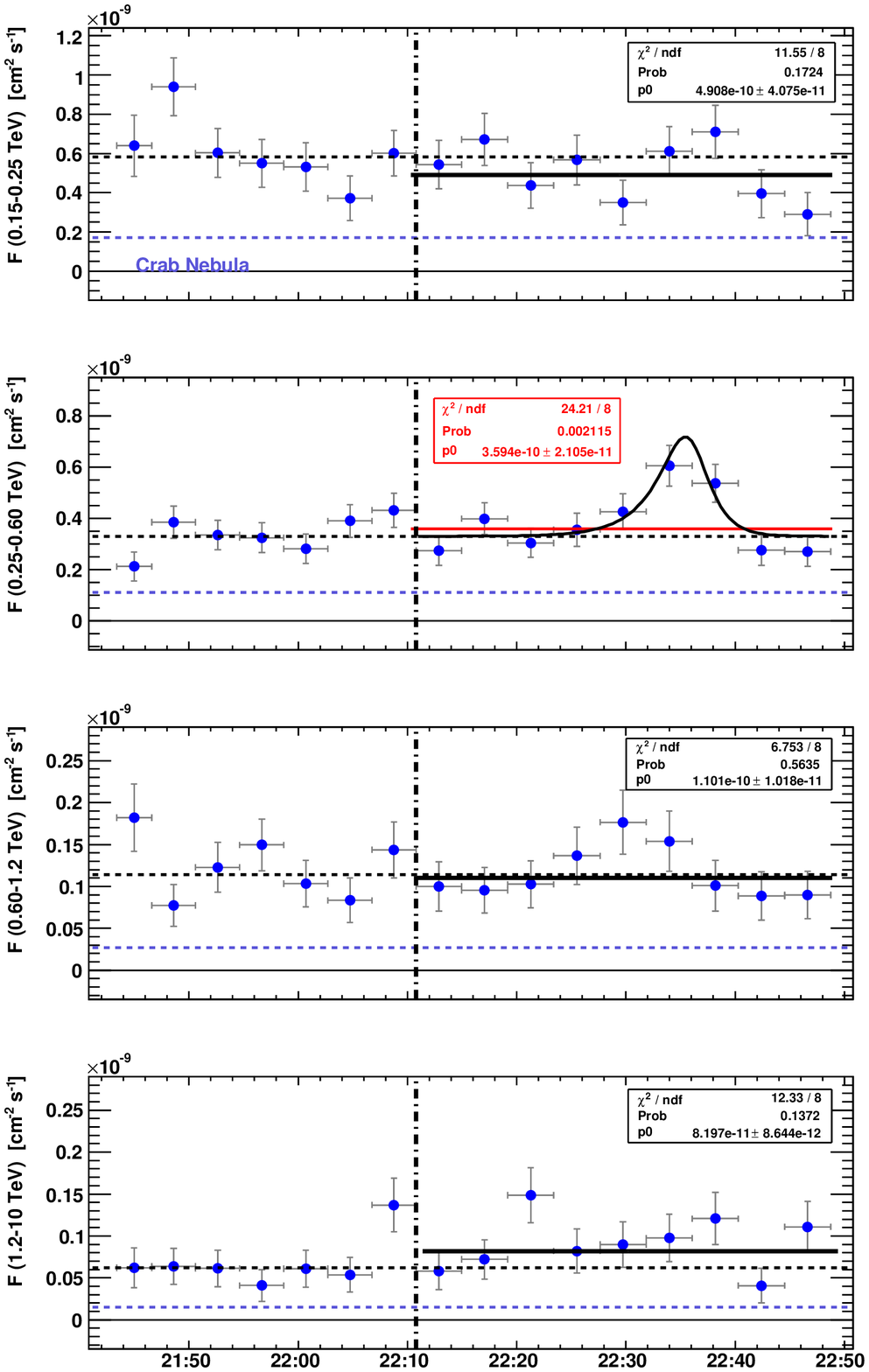}
\caption{LC for the night June 30  with a time binning of 4 minutes, 
and separated in different energy bands, from the top to the bottom, 
0.15-0.25 \TeV, 0.25-0.6 \TeV, 0.6-1.2 \TeV, 1.2-10 \TeV.
The vertical bars denote 1$\sigma$ statistical 
uncertainties. For comparison, the Crab emission is also shown as a lilac dotted horizontal line. 
The vertical dot-dashed line divides the data into 'stable' (i.e., pre-burst) and 'variable' (i.e., 
in-burst) emission emission. The horizontal black dashed line represents the average of the 'stable' 
emission.
The 'variable' (in-burst) of all energy ranges were fit with a constant line. The results of 
the fits are given in the insets. The constant line fit on the energy range 0.25-0.6 \TeV\
was not satisfactory (see inset in figure); yet this LC could be fit with the flare model 
described by equation  \ref{eq_fitflare} (see table \ref{fitflareresults_F0701_ERanges} for 
the resulting parameters).
\label{LCSingle_MultiEnergyRanges_F0701}}
\end{figure}

Because of the steeply falling spectra, the low-energy events dominate the LCs shown in Fig. 
\ref{LCSingle}, and tend to hide any higher-energy features. We therefore split the data into 
four distinct energy ranges: 0.15-0.25 \TeV, 0.25-0.6 \TeV, 0.6-1.2 \TeV\ and 1.2-10 \TeV. 
The corresponding LCs for the night June 30 are shown in Fig. \ref{LCSingle_MultiEnergyRanges_F0701}. 
Due to the reduced photon statistics, 
we increased the time binning from 2 minutes to 4 minutes. 
We found that only the energy range 0.25-0.6 \TeV\ shows a clear flux variation; 
a constant line fit gave $\chi^2/NDF=24.2/8$ ($p=2\cdot10^{-3}$). The other energy ranges 
are compatible with a constant line fit, showing only a slight overall flux 
level variation with respect to the LC in the 'stable' part. Therefore, if there is a flux variation, 
it is too small to be significantly seen in our data. Note that, among the 4 energy ranges used,
the 0.25-0.60 GeV energy range is the one with the highest sensitivity for flux variations.
The 'variable' LC from the energy range 0.25-0.60 \TeV\ was then fitted using equation 
 \ref{eq_fitflare}, fixing $t_0$ to the time of the highest point in the LC. 
The resulting parameters of the fit are reported in table \ref{fitflareresults_F0701_ERanges}; 
the rise/fall flux-doubling times are comparable to those ones obtained using the 
integrated LC above 0.15 \TeV.

\begin{figure}
\epsscale{0.7}
\plotone{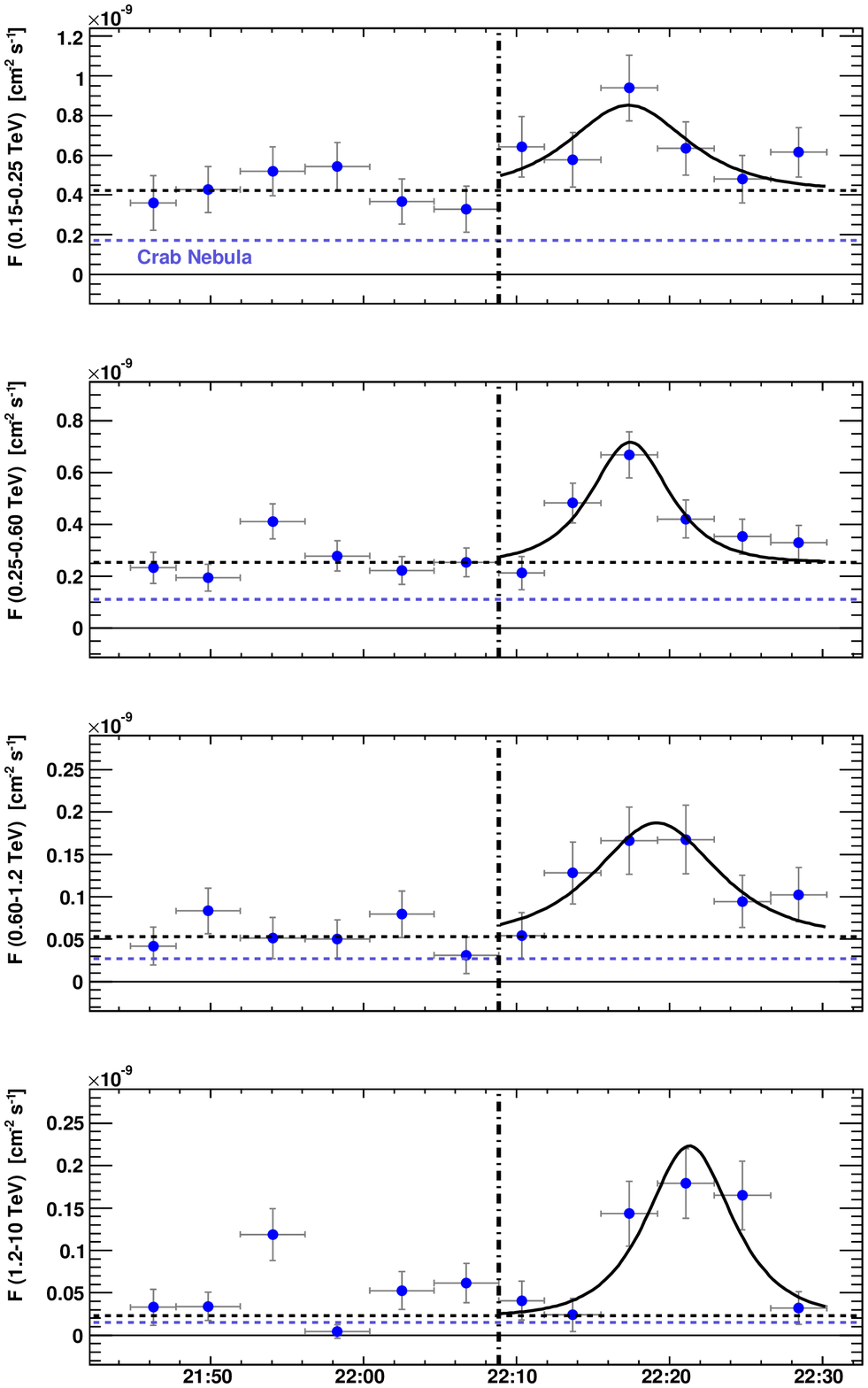}
\caption{LC for the night July 9 with a time binning of 4 minutes, 
and separated in different energy bands, from the top to the bottom, 
0.15-0.25 \TeV, 0.25-0.6 \TeV, 0.6-1.2 \TeV, 1.2-10 \TeV.
The vertical bars denote 1$\sigma$ statistical 
uncertainties. For comparison, the Crab emission is also shown as a lilac dotted horizontal line. 
The vertical dot-dashed line divides the data into 'stable' (i.e., pre-burst) and 'variable' (i.e., 
in-burst) emission emission. The horizontal black dashed line represents the average of the 'stable' 
emission.
The 'variable' (in-burst) of all energy ranges were fit with  a flare model described by 
equation \ref{eq_fitflare}, where  $c=d$ (rise=fall time). All parameters were left free 
in the fit. All light curves were considered simultaneously in the fit (combined fit).
The resulting parameters from this combined fit 
are reported in table \ref{fitflareresults_F0709_ERanges}. 
\label{LCSingle_MultiEnergyRanges_F0709}}
\end{figure}

The same exercise on the flare July 9 gave a significantly different result, as shown in Fig. 
\ref{LCSingle_MultiEnergyRanges_F0709}. The flare is visible essentially in all energy ranges. 
In order to study possible time shifts between the different energy ranges, we fit 
all LCs simultaneously (combined fit) with a flare model described by 
equation \ref{eq_fitflare}. In order to remove one degree of freedom and facilitate the 
fit procedure, we assume a symmetric flare with equal rise and fall flux-doubling times; 
that is $c=d$ in equation \ref{eq_fitflare}. The resulting parameters from this 
combined fit are shown in table \ref{fitflareresults_F0709_ERanges}.
The combined fit 
gave $\chi^2/NDF=14.0/12$ ($p=0.3$), which implies that the measured 
flare is compatible with being symmetric. The rise/fall flux-doubling time is 
about 2 minutes for all the energy ranges. 
It is interesting to note that the position of the peak of the flare for the 
different LCs seems to vary somewhat with energy. The time difference between the highest energy 
range and the lowest energy range is 239 $\pm$ 78 seconds. If instead, the energy 
range 0.25-0.6 \TeV\ is selected as the lowest energy range, 
which has a better defined flare (and thus a better 
determination of the peak position), the time difference is 232 $\pm$54 seconds. 

\begin{figure}
\epsscale{0.7}
\plotone{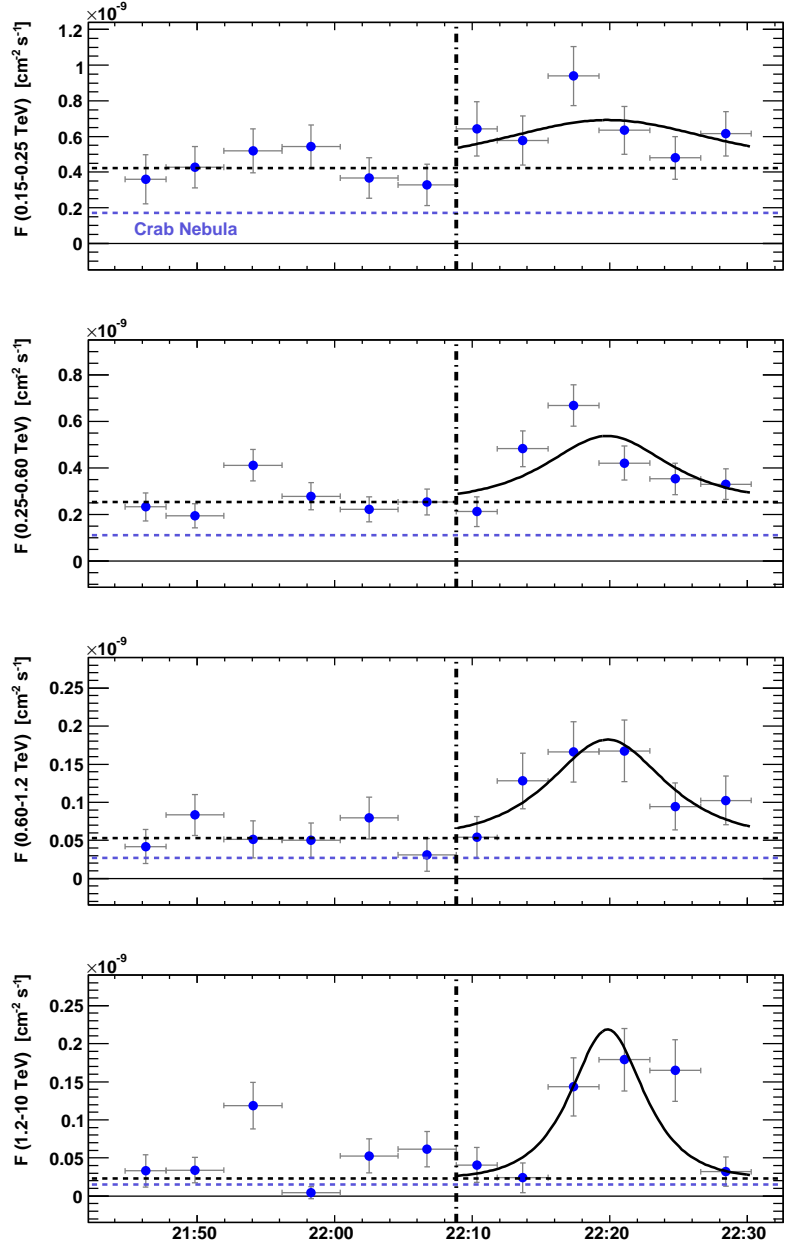}
\caption{Same as in figure \ref{LCSingle_MultiEnergyRanges_F0709}, but with a 
common T0 (which was also fit) for all LCs.  
The resulting parameters from this combined fit 
are reported in table \ref{fitflareresults_F0709_ERanges_commont0}. 
\label{LCSingle_MultiEnergyRanges_F0709_commont0}}
\end{figure}

In order to evaluate the significance of this time shift, we 
performed the same fit, but this time using a common $t_0$ for all LCs. 
The resulting fits are shown in Fig. \ref{LCSingle_MultiEnergyRanges_F0709_commont0}, 
and the resulting parameters from the fit in table \ref{fitflareresults_F0709_ERanges_commont0}. 
The combined fit gave $\chi^2/NDF=26.6/15$ ($p=0.04$), which implies that such a situation 
is unlikely, and consequently that the time shift of 4 $\pm$ 1 min between 
the highest and the lowest energies is more probable. 

Investigating the reliability of the
time delay obtained from the combined fit, 
we performed a cross-correlation analysis on the LCs from July 9 
with the methodology described in section \ref{Correlation}. For this 
study we used LCs with 2 min time bins from the energy ranges 0.25-0.6 \TeV\ and 
1.2-10 \TeV\footnote{
The flare observed in the LC from the energy range 0.15-0.25 \TeV\ is 
not very well defined because of the somewhat larger measurement errors, and 
the smaller relative amplitude of the flux variation (with respect to the 'stable' emission) 
with decreasing energy.}.
The correlation coefficient and probability of correlation were computed 
after introducing time shifts of 2 min (one bin in the LCs). We obtained the 
highest values for a time lag of 4 minutes; which is consistent with the 
results from the combined fit shown above.

We want to point out that this is the first time that a possible time delay between flares at 
different energies is observed at \vhe\ \gray\ energies, although such time lags have been 
detected for some \TeV\ blazars at \xray\ frequencies, viz. Mrk 421 \citep{Ravasio2004} and  
PKS 2155-304 \citep{Zhang2006a, Zhang2006b}. If the observed VHE time lag is assumed real,
this suggests that we are 
observing the underlying dynamics of the relativistic electrons in both the synchrotron and IC 
emission, and the observation, therefore, supports SSC models.

It should be also noted that the relative amplitude of the flux variations observed in the
LC for July 9 with respect to the 
baseline emission, is significantly larger at the highest energies. 
This can be seen from the ratio $b/a$, where $a$ and $b$ are 
the coefficients in eq. \ref{eq_fitflare} describing, respectively, the baseline and amplitude 
of the flare (see table \ref{fitflareresults_F0709_ERanges}): $b/a=3.6\pm1.0$ and 17$\pm$4 for, 
respectively, the \mbox{0.25-0.6 \TeV} and \mbox{1.2-10 \TeV} bands. The July 9 LC also shows 
some significant flux variation in its 'stable' part: in the highest-energy band, where activity 
is most conspicuous, a constant fit gives a $\chi^2/NDF=20.6/5$ ($p=9.6\cdot10^{-4}$). 

In summary, during the 2005 MAGIC observations of Mrk 501 we detected variability at \vhe\ 
frequencies with flux-doubling times down to 2 minutes. This is about 50 times faster than the shortest 
previously observed variability-times at VHE frequencies for Mrk501 \citep{Hayashida1998,Quinn1999,AharonianMrk501-1999a,
Mrk501CAT}, and about 5 times shorter than the shortest observed variability for Mrk 421 
\citep{gaidos1996}. The above presented flux-variations are among the shortest ever observed in blazars,
see also \citet{HESSPKS2155}. It is interesting to note that the Mrk 501 flux-doubling rise times observed by 
\magic\ in the \vhe\ range are rather comparable to the shortest variability times observed at \xray\ 
frequencies which were reported by \citet{XueCui2005}: 
a flare with a total duration of 15 minutes with a flux variation of 30\%. 
The authors, however, reported the presence of substructures, which point to the existence of 
variability on timescales shorter than 15 minutes. It is worth mentioning that for both 
\xray\ and \gray\, the shortest flux varitions occurred 
when the source was not in an exceptionally high emission state.

\subsection{Quantification of the variability}
\label{Quantificationvariability}

Mrk 501 has shown energy-dependent flux variations throughout the entire \magic\ observational campaign. 
We followed the description given in \citet{Vaughan2003} to quantify the flux variability by means of 
the fractional variability parameter \Fvar, as a function of energy. In order to account for the 
individual flux measurement errors ($\sigma_{err,i}$), we used the 'excess variance' \citep{Nandra1997,
Edelson2002} as an estimator of the intrinsic source variance. This is the variance after subtracting 
the expected contribution from measurement errors. For a given energy range, the \Fvar\ is calculated as
\begin{equation}
\label{form_nva}
\Fvar = \sqrt{\frac{S^2 - <\sigma_{err}^2 >}{<F_{\gamma}>^2}}
\end{equation}
where  $<F_{\gamma}>$ is the mean photon flux, $S$ the standard deviation of the $N$ flux points, and 
$<\sigma_{err}^2>$ the average mean square error, all determined for a
  given energy bin.
The uncertainty on \Fvar\ is estimated according to:
\begin{equation}
\label{form_nva_err}
\Delta \Fvar = \sqrt{\Big(\sqrt{\frac{1}{2 N}} \cdot \frac{<\sigma_{err}^2>}{<F_{\gamma}>^2 \cdot \Fvar}\Big)^2 + 
  \Big(\sqrt{\frac{<\sigma_{err}^2>}{N}}\frac{1}{<F_{\gamma}>}\Big)^2}
\end{equation}

\begin{figure}
\epsscale{1}
\plottwo{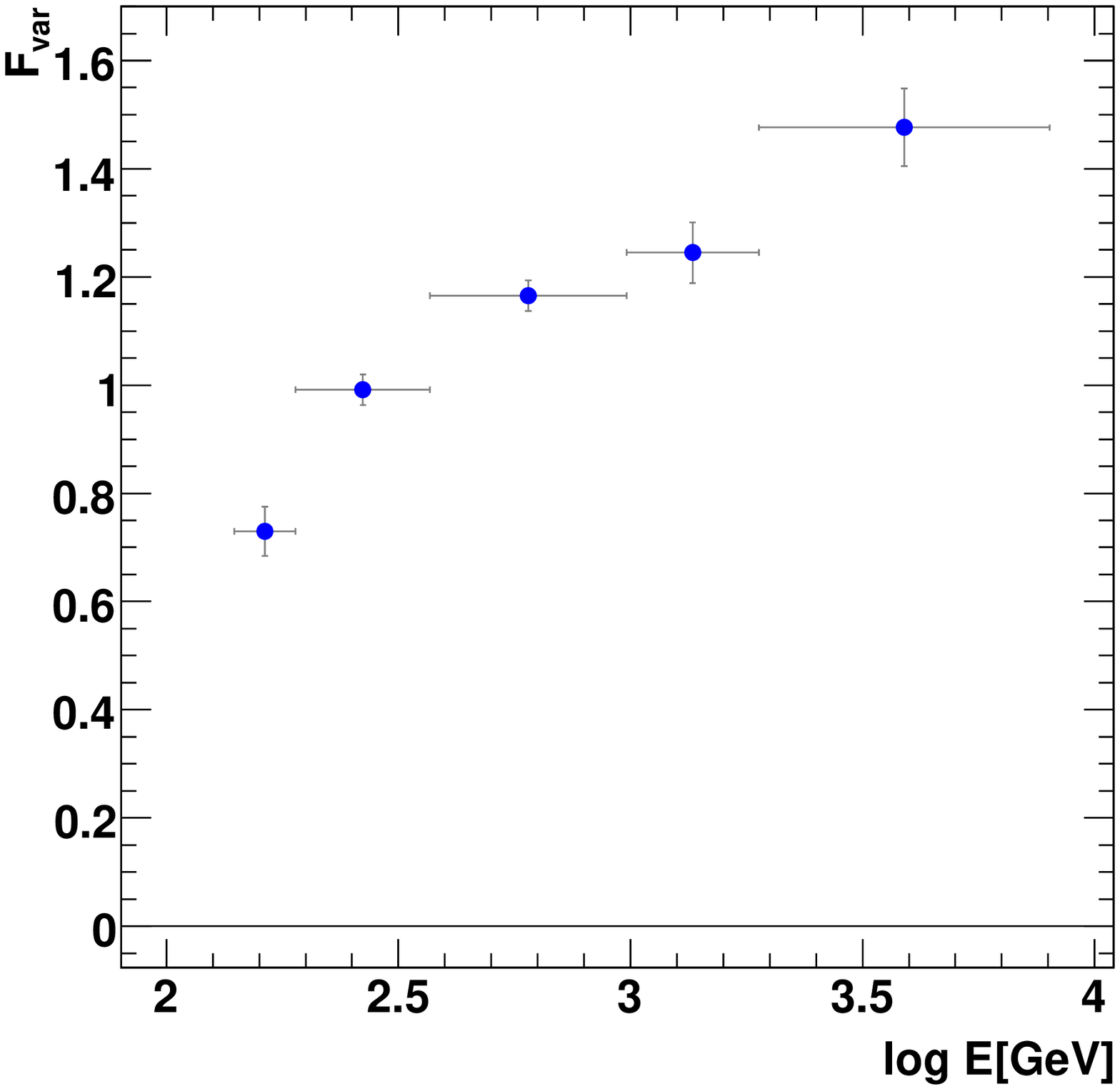}
{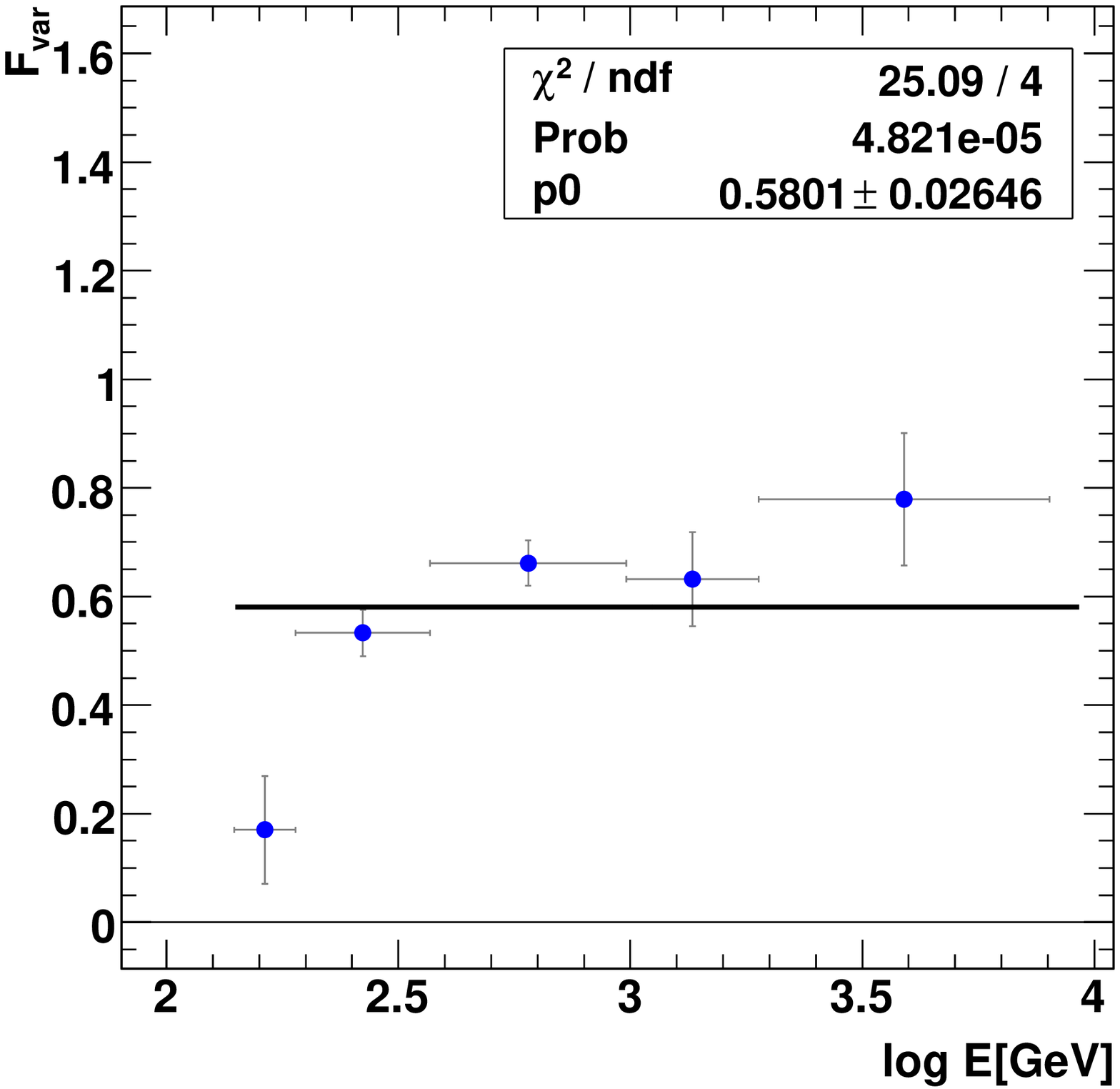}
\caption{
Fractional variability parameter as derived for 5 energy bins. Vertical bars denote 1$\sigma$ 
uncertainties, horizontal bars indicate the width of each energy bin. The left plot includes 
all data. The right plot includes all but the June 30 and July 9 data. The black horizontal 
line in the right panel results from a constant fit to the data points (see inset for fit 
parameters). 
\label{fig_nva}}
\end{figure}

Fig. \ref{fig_nva} shows the derived \Fvar\ values for 5 logarithmic energy bins, spanning from \mbox{0.14 \TeV} 
to 8 \TeV \footnote{
The \Fvar\ is not meaningful below 0.14 \TeV\ (i.e. below the threshold energy of the instrument)
because the flux errors are rather large, which makes $<\sigma_{err}^2> \sim S^2$.}.
The left-hand plot includes 
all data, while the right-hand plot includes all data except for the two active nights. The result indicates a 
larger-amplitude flux variation at higher energies, which is clearly discernible even when the active-state data are 
excluded (the null probability being $p$$\sim$$10^{-5}$).

\begin{figure}
\epsscale{1}
\plottwo{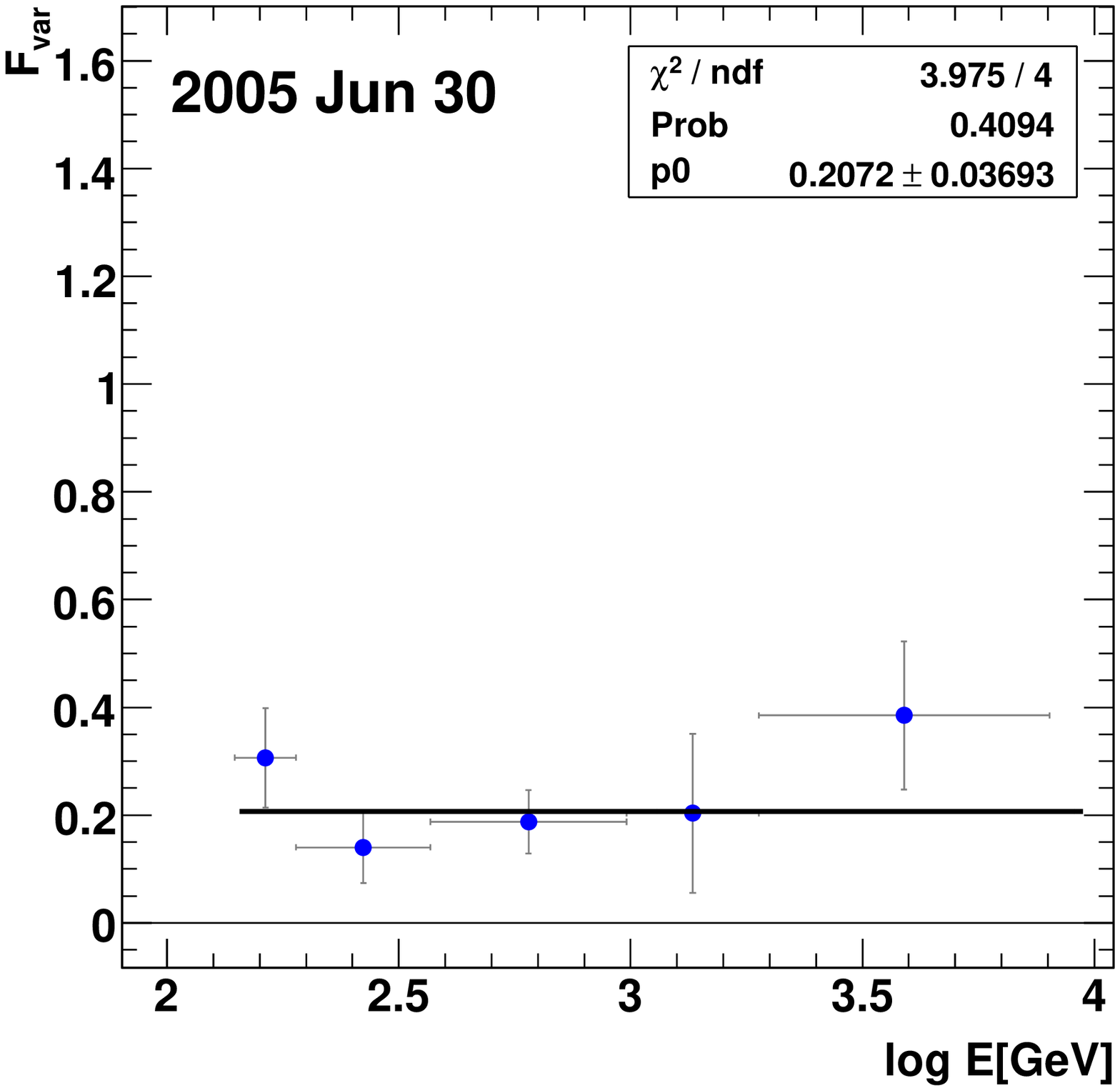}
{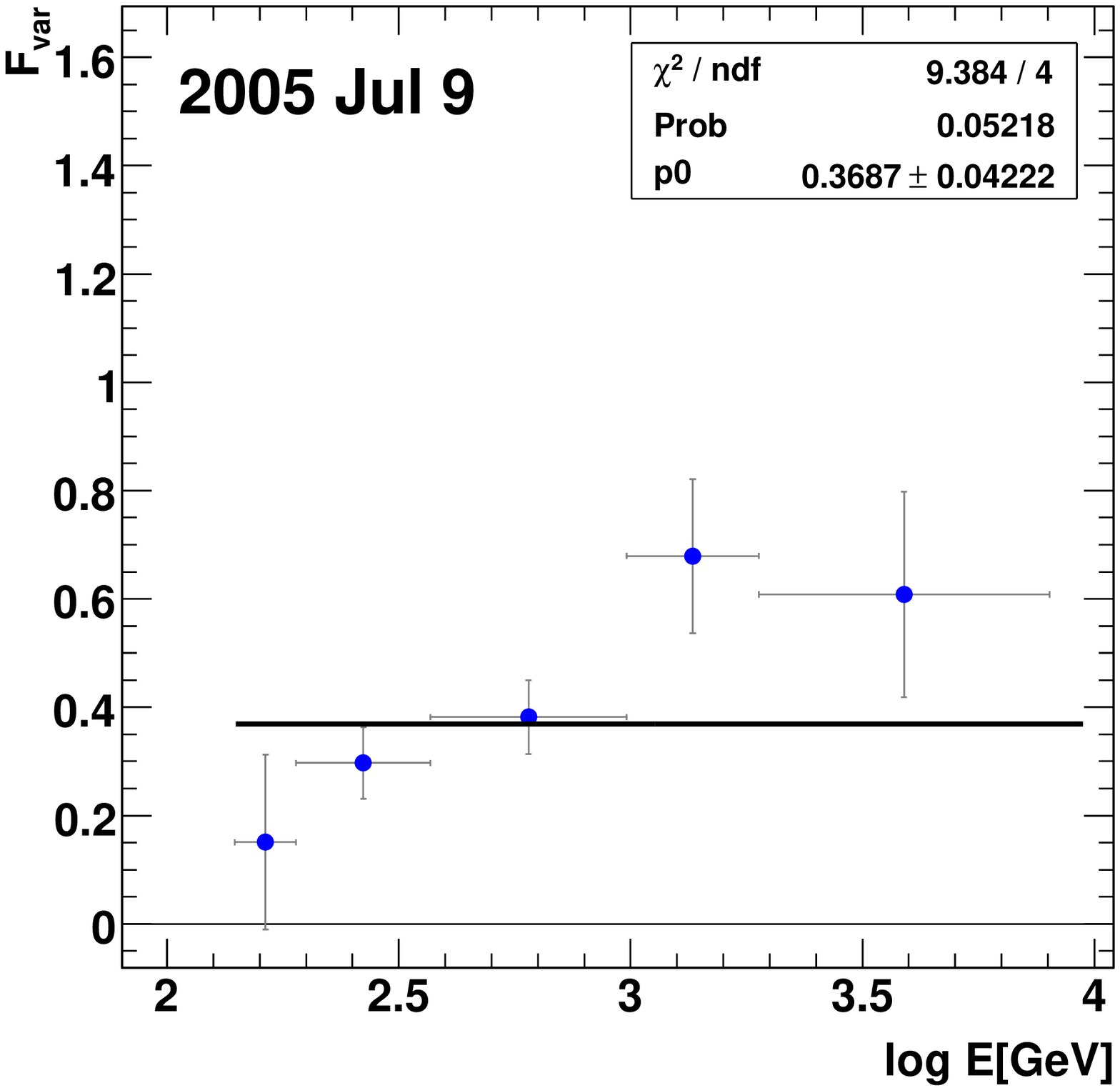}
\caption{Fractional variability parameter as computed for the nights of June 30 and July 9. Vertical 
bars denote 1$\sigma$ uncertainties, horizontal bars indicate the width of each energy bin. The black 
horizontal lines are the result of a constant fit to the data points (see inset for fit parameters).
\label{fig_nva_singlenights}}
\end{figure}

Fig. \ref{fig_nva_singlenights} shows the \Fvar\ values as derived for the individual active nights of June 
30 and July 9. The flux variability in these nights is smaller than for the entire observational campaign, 
as one would expect from simple inspection of Fig. \ref{OverallLightCurve}. In the night of June 30, \Fvar\ 
does not increase significantly with energy. In contrast to June 30, there is a clear increase of \Fvar\ with 
energy in the night of July 9 - in spite of the larger error bars coming from a shorter observation time and 
a lower mean flux. 

In summary, during the year 2005 \magic\ observations, the \vhe\ \gray\ flux variability 
of Mrk 501 was found to significantly increase with energy, on time scales from months to less than an hour. 
A similar effect (on timescales $\magcir$1/2 hr) was also detected in X-rays in 1997, 1998, and 2000 (\citet{
gliozzi2006}, based on RXTE data). Another \xray\ evidence was found for the \TeV\ blazar: PKS 2155-304 \citep{
Zhang2006b}. The largest \xray\ \Fvar\ value in X-rays for Mrk 501 was $\sim$0.6-0.7 in the highest energy 
bin at 10-20 \keV, and it was found in June 1998 and July-September 2000 (\citet{gliozzi2006}). In 1997 however, 
despite Mrk 501 showed the highest \xray\ (2-10 \keV) fluxes from the last ten years, 
the largest observed \Fvar\ value in \xrays\  was only $\sim$0.4. 
It is interesting to note that, in the \vhe\ \gray\ range, we observe a maximum \Fvar\ of $\magcir$0.6, which 
increases to $\magcir$1.2 when the two active nights are included. This suggests that Mrk 501 is more variable 
in VHE \grays\ than in \xrays.

\section{\vhe\ Spectra}
\label{Sed}

The differential photon spectra of \mbox{Mrk 501} were parameterized with a simple power-law (PL) 
function:
\begin{equation}
\label{eq_powerlaw}
{dF \over dE} = K_0 \cdot \Big(\frac{E}{0.3 \TeV}\Big)^{-a}
\end{equation}
where $K_0$ is a normalization factor and $a$ the photon index. 
The MAGIC sensitivity permits to derive VHE spectra of Mrk 501 
on a daily basis, independent of its flux level, a real asset for unbiased precision 
studies of blazars. The PL spectral parameters for each single night are reported in  
Table \ref{SummaryTable}. 

During our observations the \vhe\ emission of \mbox{Mrk 501} showed a very dynamic behaviour, with 
significant spectral variability on a timescale of days (see Sect. \ref{SedCorrelation} and \ref{Sed4}). 
Nevertheless, most of the data 
are well described by the simple PL function of eq. 6. This does not hold for the two flaring nights of June 30 and 
July 9 which are therefore discussed separately in Sect. \ref{SEDBigFlares} and \ref{Sedintradayvariations}.

\subsection{Spectral index vs. flux}
\label{SedCorrelation}

\begin{figure}
\epsscale{1}
\plotone{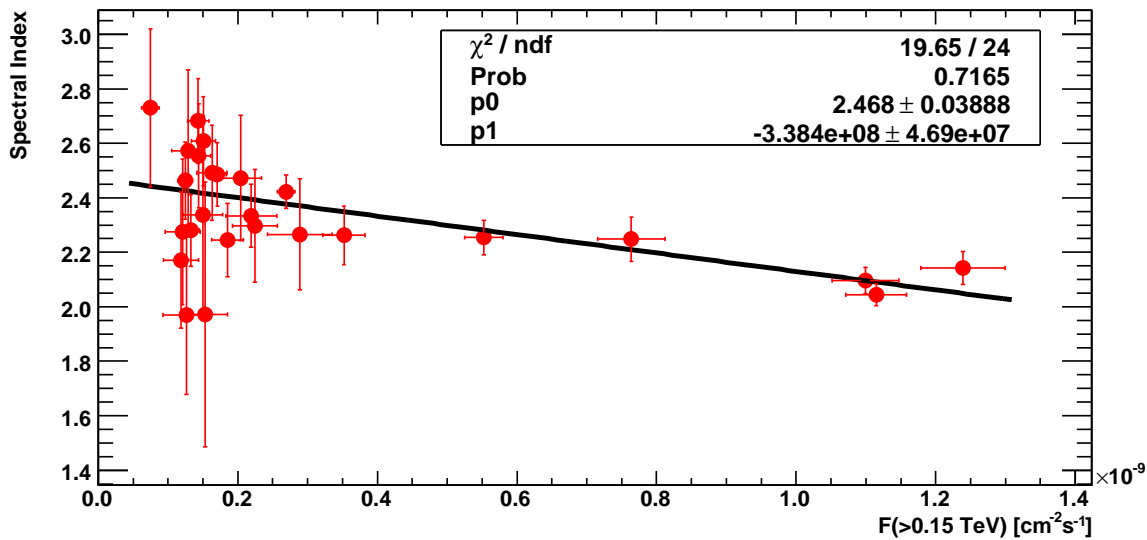}
\caption{Correlation between spectral shape and integrated flux above 0.15 \TeV. Each point 
denotes a single night of observation. The error bars denote 1$\sigma$ uncertainties. The 
June 30 and July 9 data were split chronologically into two data sets each, corresponding 
to the pre-flare ('stable') and in-flare ('variable') emission shown in figure \ref{LCSingle}.  
\label{CorrIndexFlux}}
\end{figure}

The Mrk 501 \vhe\ spectrum was measured on a night-by-night basis, which allowed investigating possible 
correlations between the PL spectral index and intensity, as it is shown in Fig. \ref{CorrIndexFlux}. 
The data of June 30 and 
July 9 were again split into the 'stable' and 'variable' part (see Fig. \ref{LCSingle}) and consequently, Fig. 
\ref{CorrIndexFlux} contains 26 instead of 24 points. The data points are well described by a linear fit; but a 
constant is clearly excluded ($\chi^2/NDF=74.7/25$, i.e. $p=7.4\cdot10^{-7}$). On average, the \mbox{Mrk 501} 
spectrum hardens when the emission increases. Such a correlation was already reported by \citet{Pian1998,
Tavecchio2001}, although on substantially longer timescales.

\begin {figure} 
\epsscale{0.5}
\plotone{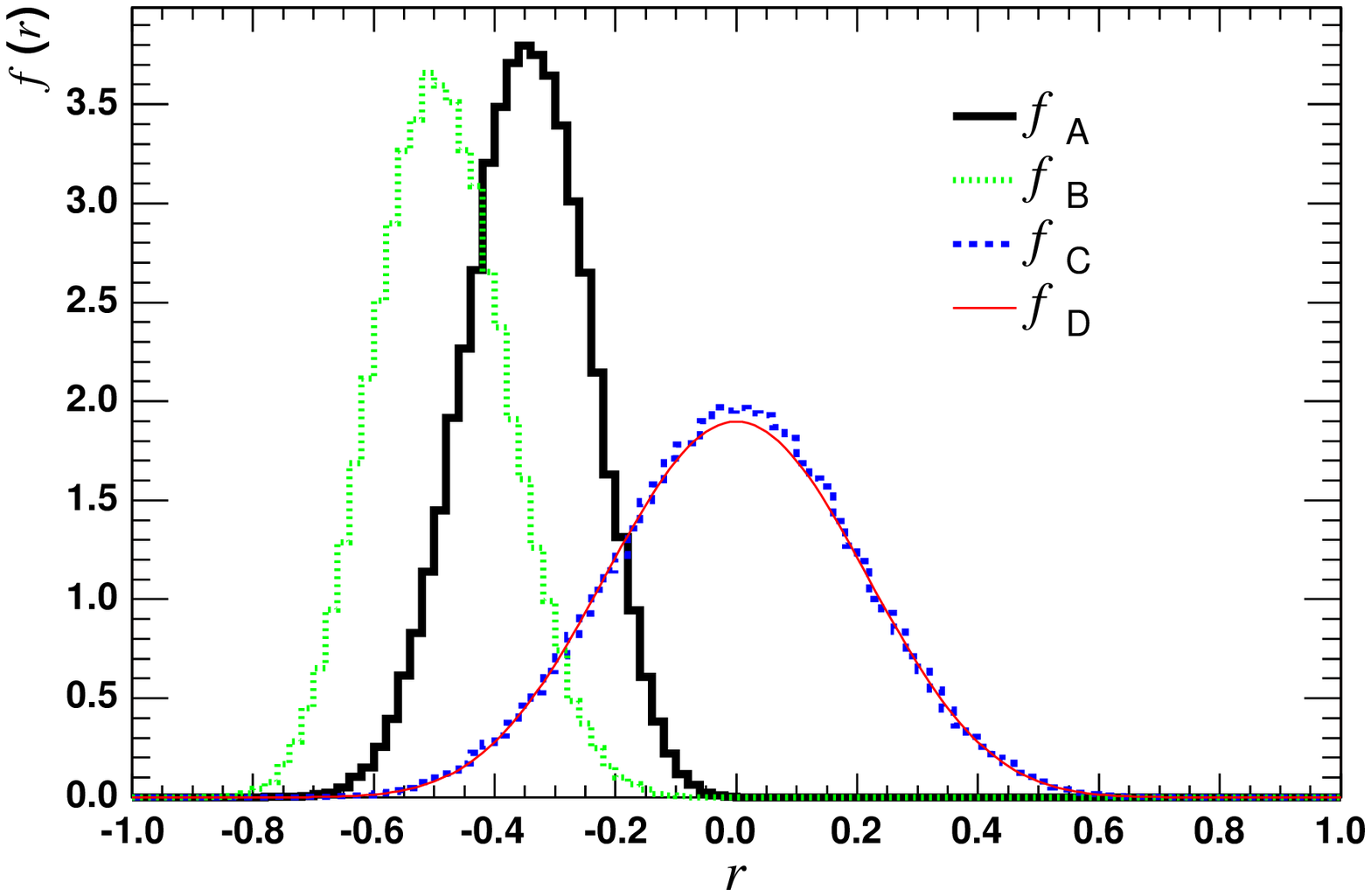}
\caption{Probability density
functions of the correlation coefficient between spectral index and \gray\ fluxes: 
$f_A$ data, $f_B$ perfectly correlated case, $f_C$ uncorrelated case, 
and $f_D$ analytical solution  in the uncorrelated case.
See text for further details.
\label {CorrIndexFluxCalc}}
\end {figure}

In order to test the ansatz of a linear correlation in spectral index versus \gray\ flux (see Fig. 
\ref{CorrIndexFlux}), we applied a correlation analysis as described in Sect. \ref{Correlation}. The linear fit 
indicates anti-correlation ($\chi^2/{\rm ndf} = 19.65/24$, $p$$=$$0.72$). 
The correlation factor is -0.48, and the corresponding probability 
distributions are shown in Fig. \ref{CorrIndexFluxCalc}. We find the probability $P_{\alpha,\gamma}(f_A,f_B) = 
0.35$, to be higher than $P_{\alpha,\gamma}(f_A,f_C) = 0.12$, which supports a modest anti-correlation. 

Similar, the probability for the first scenario to be true and for the second to be false is $P_{{\alpha},
\gamma}(f_A,f_B)(1 - P_{{\alpha},\gamma}(f_A,f_C)) = 0.30$, while the probability for the opposite case is $(1 - 
P_{{\alpha},\gamma}(f_A,f_B)) P_{{\alpha},\gamma}(f_A,f_C) = 0.08$. In conclusion, this correlation study 
indicates a spectral hardening with increased flux.

\subsection{Spectra at different flux levels}
\label{Sed4}

\begin{figure}
\epsscale{0.5}
\plotone{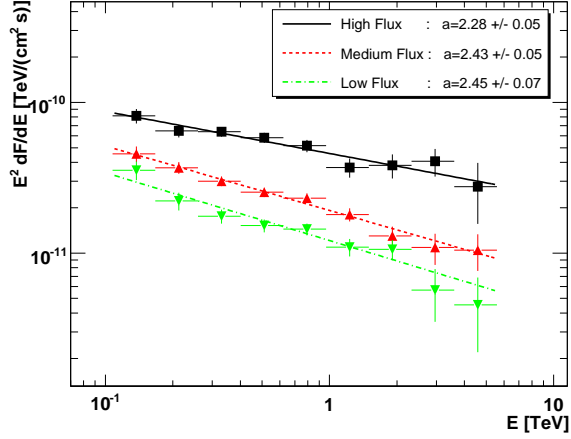}
\caption{Measured spectra of \mbox{Mrk 501} for three different $>$0.15 \TeV\ flux levels: low (green down  
triangles), medium (red up triangles), and high (black squares). Vertical bars denote 1$\sigma$ 
uncertainties, horizontal bars denote energy bins. Lines show power-law best fits. See table \ref{FluxlevelsTable}
for fit parameters.
\label{Sed4groups}}
\end{figure}

In order to investigate the spectra at different flux levels, the diurnal data were combined into three groups, 
depending on whether the integral flux above 0.15 \TeV, $F_{0.15\,{\rm TeV}}$ (measured in {\it Crab Units} (\cu)), 
was low ($F_{0.15\,{\rm TeV}}$$<$$0.5~\cu$), medium ($0.5~\cu$$<$$F_{0.15\,{\rm TeV}}$$< 1.0~\cu$), or high ($1.0~
\cu$$<$$F_{0.15\,{\rm TeV}}$). Based on the chosen flux limits the low-, medium- and high-flux data sets consist of 
12, 8 and 2 nights, respectively (see table \ref{SummaryTable} for detailed statistics). The data from June 30 and 
July 9 will be discussed in Sect. \ref{SEDBigFlares} and are not included in the analysis here. The differential 
photon spectrum for all three flux regions together with the PL fit results are shown in Fig. \ref{Sed4groups}; 
the fit parameters are also listed in table \ref{FluxlevelsTable}. Even with such a simple 
parameterization, our data do suggest a spectral hardening with increasing flux. The results of this analysis are 
consistent with the trend seen in Fig. \ref{CorrIndexFlux} and discussed in sect. \ref{SedCorrelation}.

The HEGRA CT system measured the spectra of Mrk 501 in 1998-1999, 
at a time when its flux level was substantially below the one of the 
Crab ($\sim$1/3 \cu). The observation covered 122 hours \citep{AharonianMrk501-2001}.
The reported spectrum could be fitted with a PL in the energy range  0.5-10 \TeV,
giving a spectral index of 2.76 $\pm$ 0.08\footnote{
An exponential cutoff of $\sim$5 \TeV\ was suggested in \citet {AharonianMrk501-2001},
although the experimental data are perfectly compatible with both  hypotheses, the simple
PL $(\chi^2/NDF=12.9/14; P = 0.53)$, and the
PL with exponential cutoff $(\chi^2/NDF=9.2/13; P = 76)$.}.
It is interesting to note that this spectral index is slightly softer  than the 2.45 $\pm$ 0.07
we obtained for the low flux (17 hours of observation, $\sim$0.4 \cu) in  the energy range 0.1-6 \TeV.
This spectral shape difference might be caused by a possible  softening of the spectra above 1-2 \TeV.
Note that the fits to the spectra measured by \magic\ are mostly  constrained by the points
below 2 \TeV\ (due to low photon statistics at the highest energies),  while the fits to
the spectra measured by HEGRA are mostly constrained by the points  above 1 \TeV.
Certainly, a factor contributing to the softening of the spectra at  the highest energies
is the \gray\ extinction due to pair production by interaction with  the 
Extragalactic Background Light (EBL). According to
\citet{EBLKneiske}, the attenuation of \grays\ coming from Mrk 501 is  $\sim$30\%
at 1 \TeV\ and $\sim$50\% at 10 \TeV, while it is $\leq$15\% for  energies below 0.5 \TeV.

\subsection{Spectra during active nights}
\label{SEDBigFlares}
\begin{figure}
\epsscale{1}
\plottwo{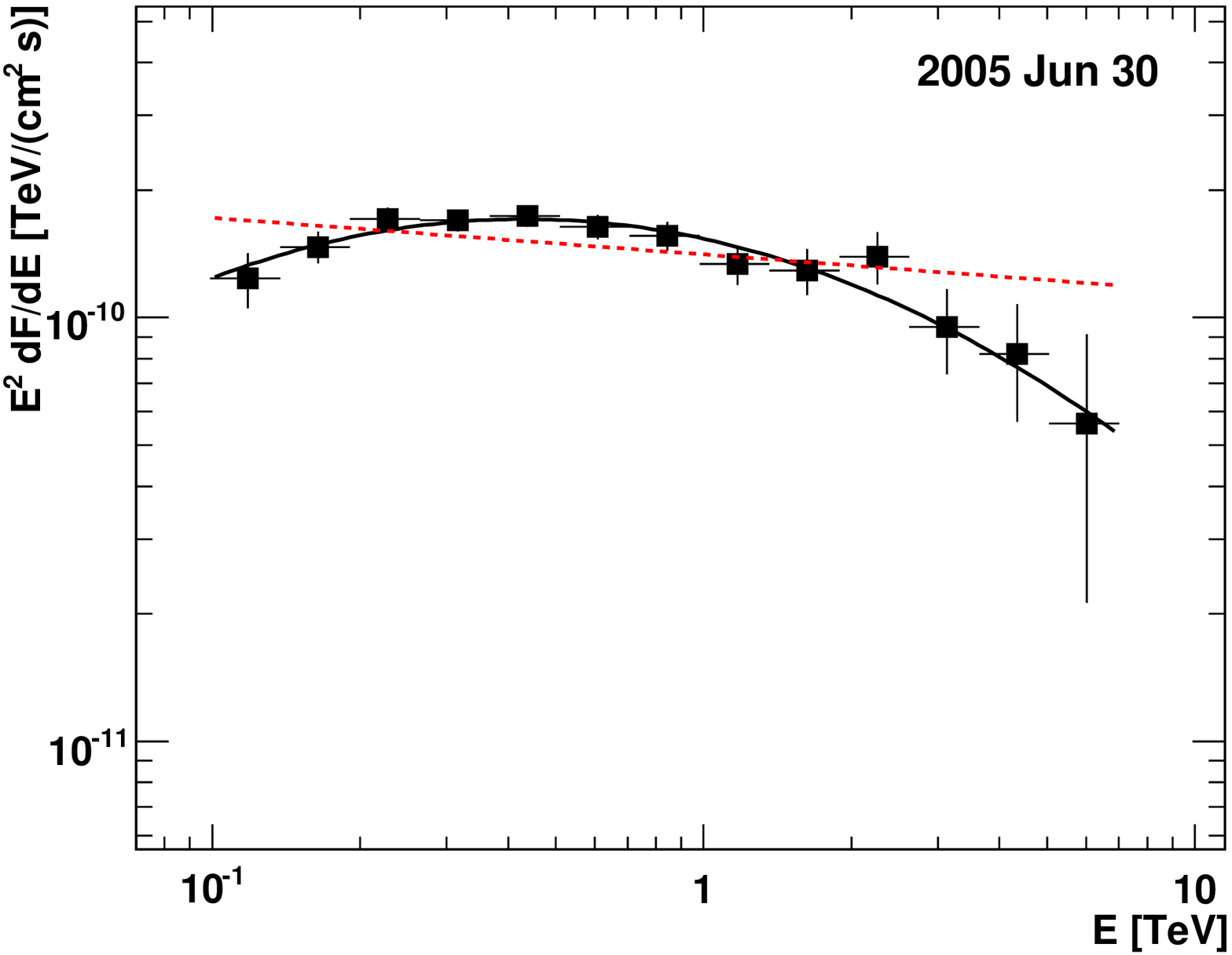}
{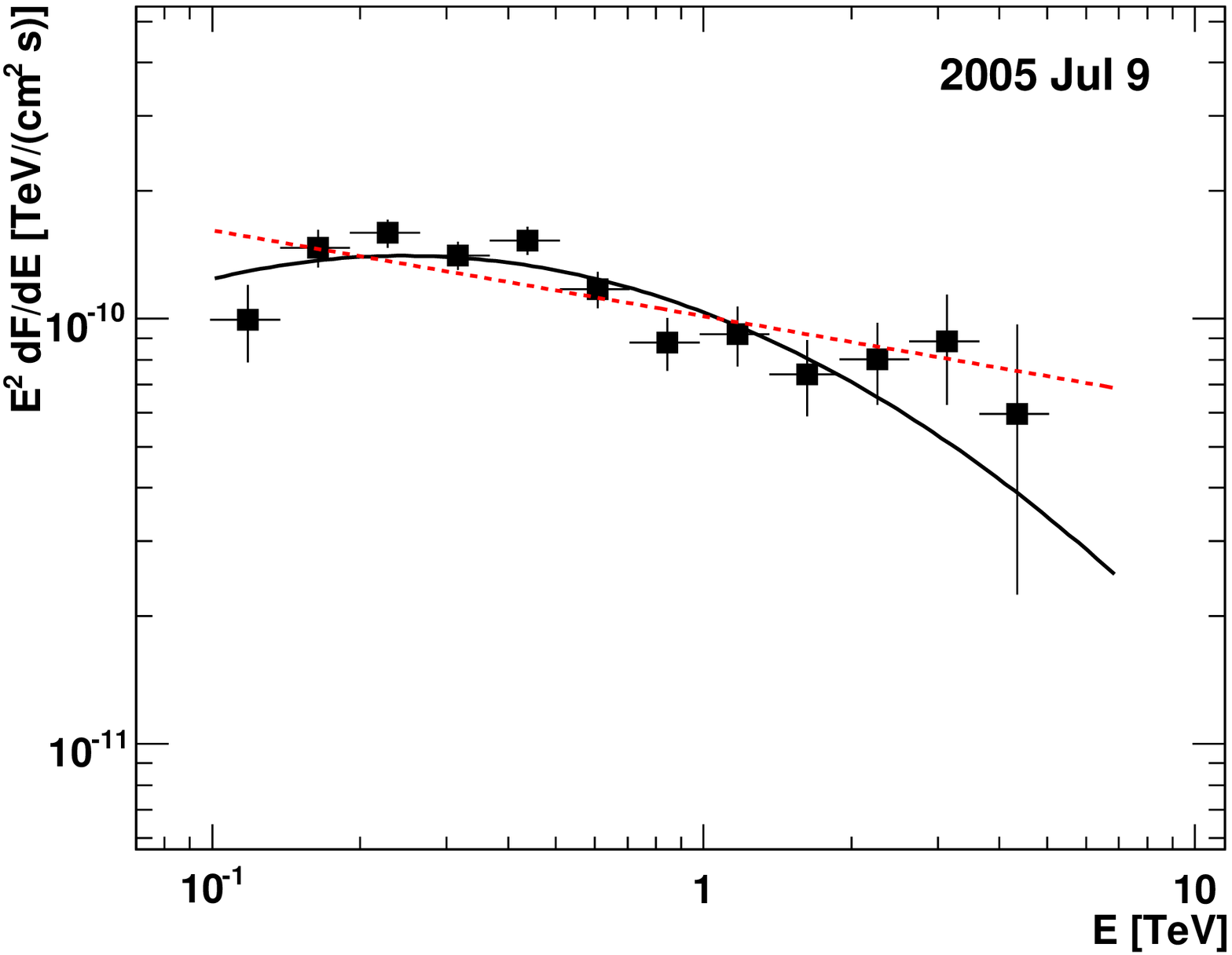}
\caption{Measured spectra for the nights of June 30 and July 9 when \mbox{Mrk 501} flared. 
Bars as in Fig. \ref{Sed4groups}. Spectral fits are a power-law (red dashed line; see eq. \ref{eq_powerlaw}) 
and a log-parabolic function (black solid line; see eq. \ref{eq_powerlawEDepInd}). See table \ref{FlareFitComparisonTable}
for fit parameters.
\label{fig_BigFlaresSED}} 
\end{figure}

The differential photon spectra from June 30 and July 9 were fitted with the simple PL described in eq. 
\ref{eq_powerlaw} as well as with the log-parabolic function: 
\begin{equation}
\label{eq_powerlawEDepInd}
{dF \over dE} ~=~ K_0 \cdot \Big(\frac{E}{0.3 \,TeV}\Big)^{-a 
           - b \cdot log_{10} \big(\frac{E}{0.3\, TeV}\big)}\,.
\end{equation}
Here, $K_0$ is a normalization factor, $a$ is the spectral index at 0.3 TeV, and $b$ is a curvature parameter 
(for $b$$>$0 the spectrum hardens/softens at energies below/above 0.3 TeV). The log-parabola is a simple function 
to describe curved spectra and, as pointed out by \citet{massaro2004,massaro2006}, can be directly related to the
intrinsic physical processes occurring in the source. The results of both, the power law and the log-parabolic 
functions are reported in Table \ref{FlareFitComparisonTable}. Since the log-parabolic fit describes the data more 
accurately than the PL fit, this suggests that the differential photon spectra of the two flaring nights are 
curved. The spectra and their fits are shown in Fig. \ref{fig_BigFlaresSED}. Note that the peak of the 
energy spectrum is located in the covered energy range in both nights.

The peak location in a spectrum described by eq. \ref{eq_powerlawEDepInd} is given by: 
\begin{equation}
\label{eq_SED_Peak}
E_{peak} ~=~ 10^{\frac{2-a}{2b}}\cdot 0.3~\TeV
\end{equation}
with an associated uncertainty of 
\begin{equation}
\label{eq_SED_PeakError}
\Delta E_{peak} ~=~ E_{peak}\cdot\frac{ln10}{2b} \cdot \sqrt{V_{aa} + 
 V_{bb} \cdot \Big(\frac{a-2}{b}\Big)^2  - 2 V_{ab} \cdot \Big(\frac{a-2}{b}\Big)}
\end{equation}
where $V_{aa},V_{bb}$ and $V_{ab}$ are the coefficients of the covariance matrix. 
Using these equations one finds that the peak locations are 
0.43$\pm$0.06 \TeV\ and 0.25$\pm$0.07 \TeV\ for the spectra measured during June 30 and July 9, respectively. 
It should be noted that 
these spectra are not corrected for EBL absorption, and are therefore not 
intrinsic to Mrk 501. After correction for EBL absorption the spectral peaks are shifted towards higher energies 
(see Sect. \ref{EBLCorr}).

The detection of a spectral curvature and measurement of the peak location was first reported by the 
Whipple and CAT collaborations 
\citep{Samuelson1998,Mrk501CAT,Mrk501CATPiron,Mrk501CATPironMoriond} based on 1997 Mrk 501 data. However, 
in those studies the spectra were not corrected for the EBL absorption. Such correction is relevant because the 
measured curvature in those energy spectra occurred essentially above 1-2 \TeV, where current EBL models predict a 
$\geq$40\% attenuation. Below 1 \TeV, little (if at all) curvature could be seen within the quoted 
1$\sigma$ statistical errors. Therefore, we suggest that the curvature and peak location 
reported in those studies may be significantly affected (if not dominated) by the EBL absorption.
On the other hand, the spectral curvature in the \magic\ data is dominated by points below 1 \TeV, 
since the higher-energy points hardly constrain the fit due to their substantially larger error bars. The 
curvature we measure thus is affected, but not dominated by EBL absorption.

\subsection{Intra-night spectral variations}
\label{Sedintradayvariations}

\begin{figure}
\epsscale{1}
\plottwo{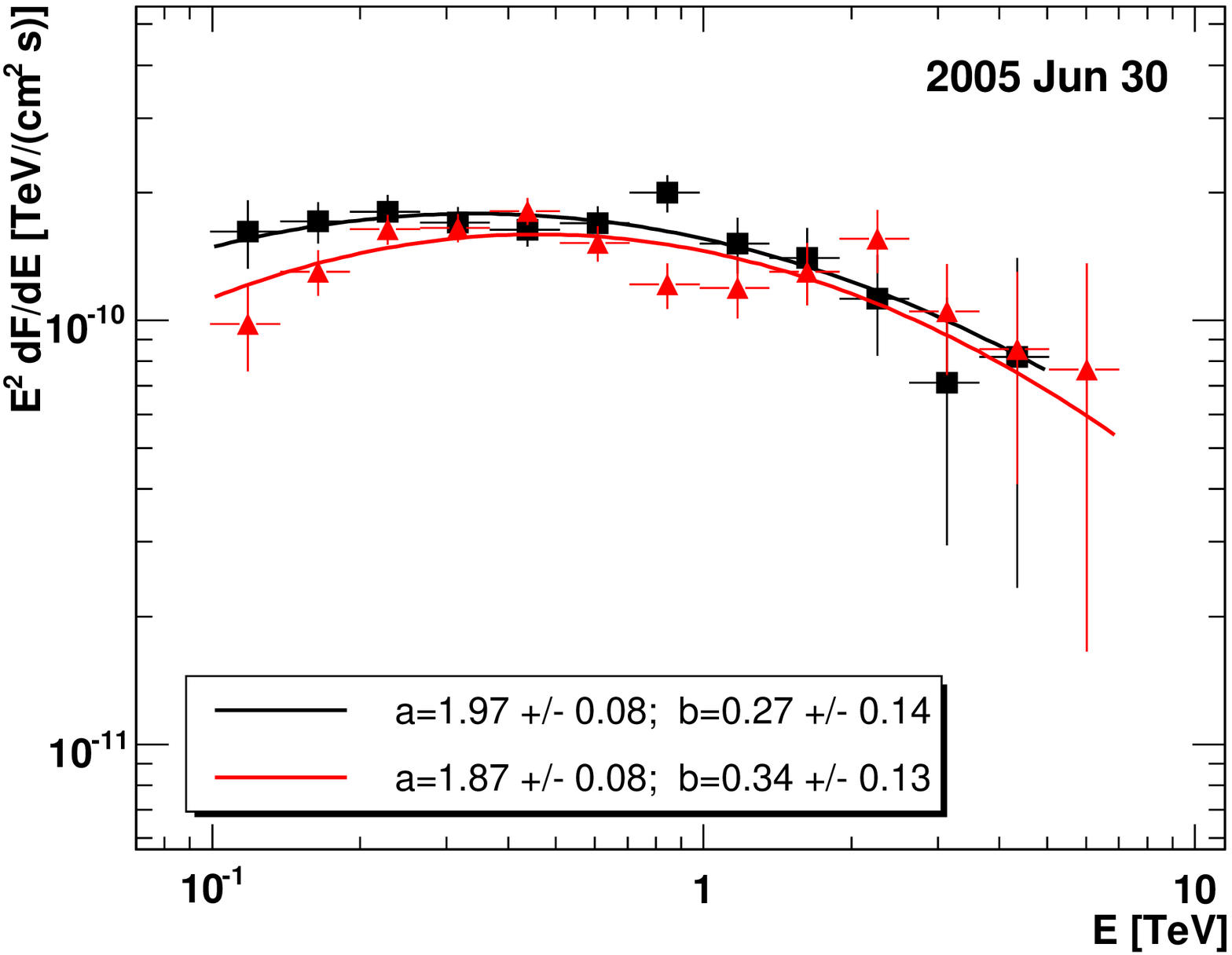}
{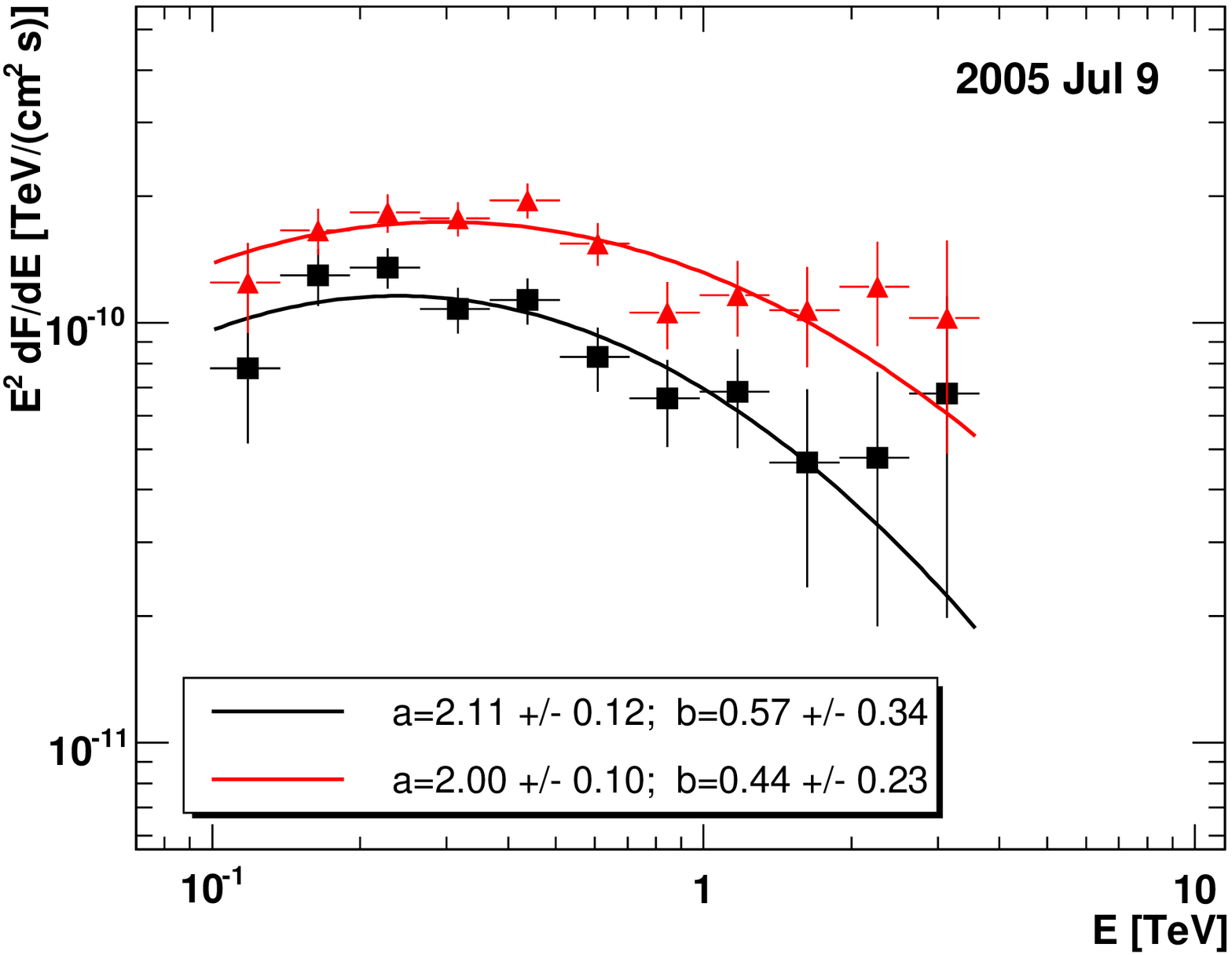}
\caption{The spectra of \mbox{Mrk 501} in the nights of June 30 and July 9 corresponding to the 
pre-burst ('stable') and in-burst ('variable') emission (see sect. \ref{Intradayvariations} 
and Fig. \ref{LCSingle}). Black squares/red triangles denote 'stable'/'variable' emission. Bars as in 
Fig. \ref{Sed4groups}. The insets show the log-parabolic fit parameters (see eq. \ref{eq_powerlawEDepInd}). 
\label{fig_Intra-day-SpectralVariations}}
\end{figure}

As discussed in section \ref{Intradayvariations}, during the active nights of June 30 and July 9 the VHE emission 
of Mrk 501 can be divided into a 'stable' (pre-burst) and 'variable' (in-burst) part. 
In order to study potential changes in the spectral 
shape, we derived the differential photon spectrum for the two parts of each  night (Fig. \ref{LCSingle}). Since 
each spectrum is based on $\leq$1/2 hr exposure this procedure certainly increases the statistical errors. The 
four spectra were fitted with the log-parabolic function in eq. \ref{eq_powerlawEDepInd}, which is preferred over 
the simple PL function (see sect. \ref{SEDBigFlares}). The results of the fit, as well as other relevant 
information (e.g., net observing time, significance of the signal, goodness of fit) are reported in Table 
\ref{FlareNightsSplitSEDTable}. The spectra and the corresponding fits are plotted in Fig. 
\ref{fig_Intra-day-SpectralVariations}. In both nights, there is marginal ($~1\sigma$) evidence for a spectral
hardening during the flare.

\begin{figure}
\epsscale{1}
\plottwo{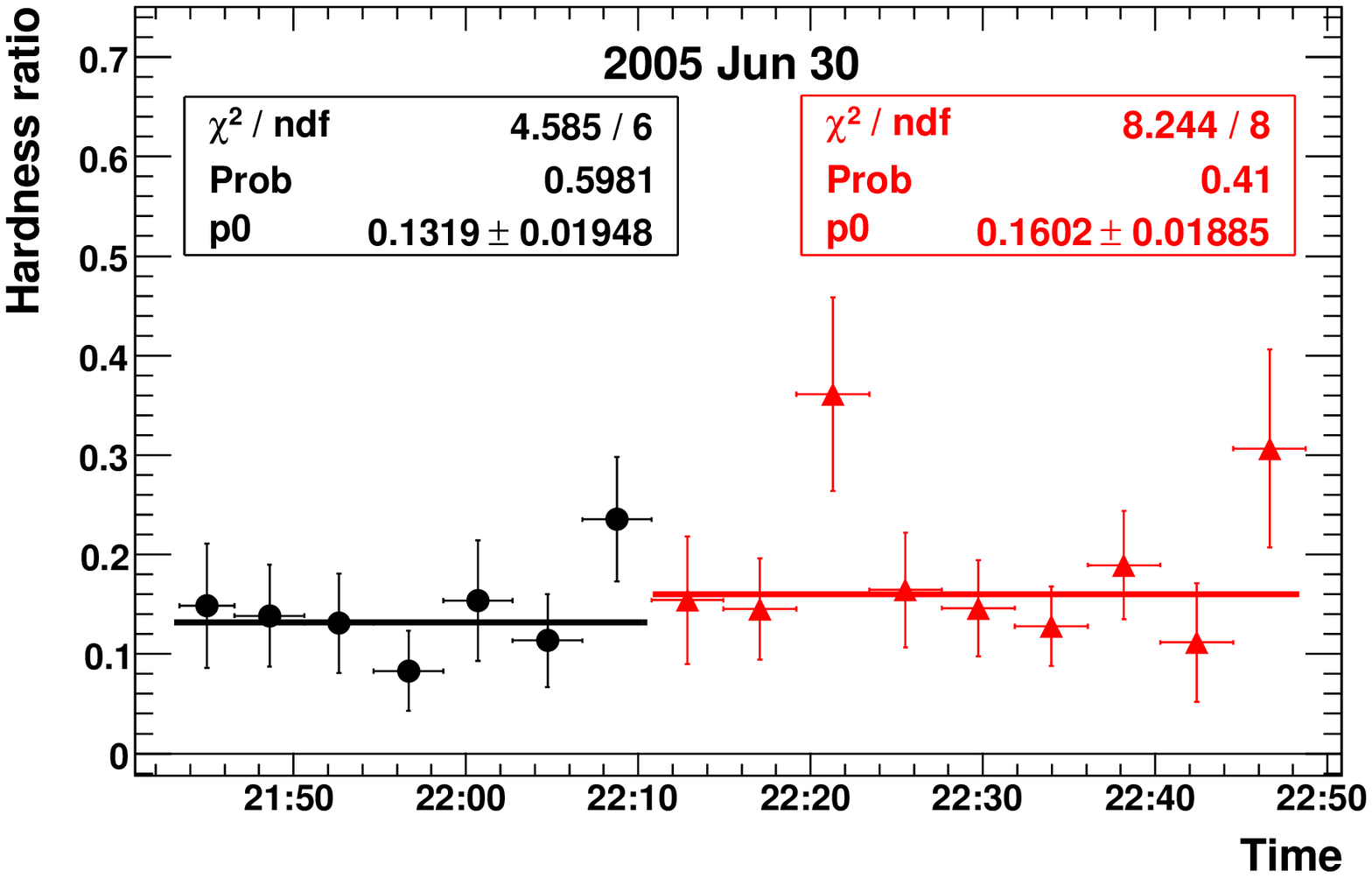}
{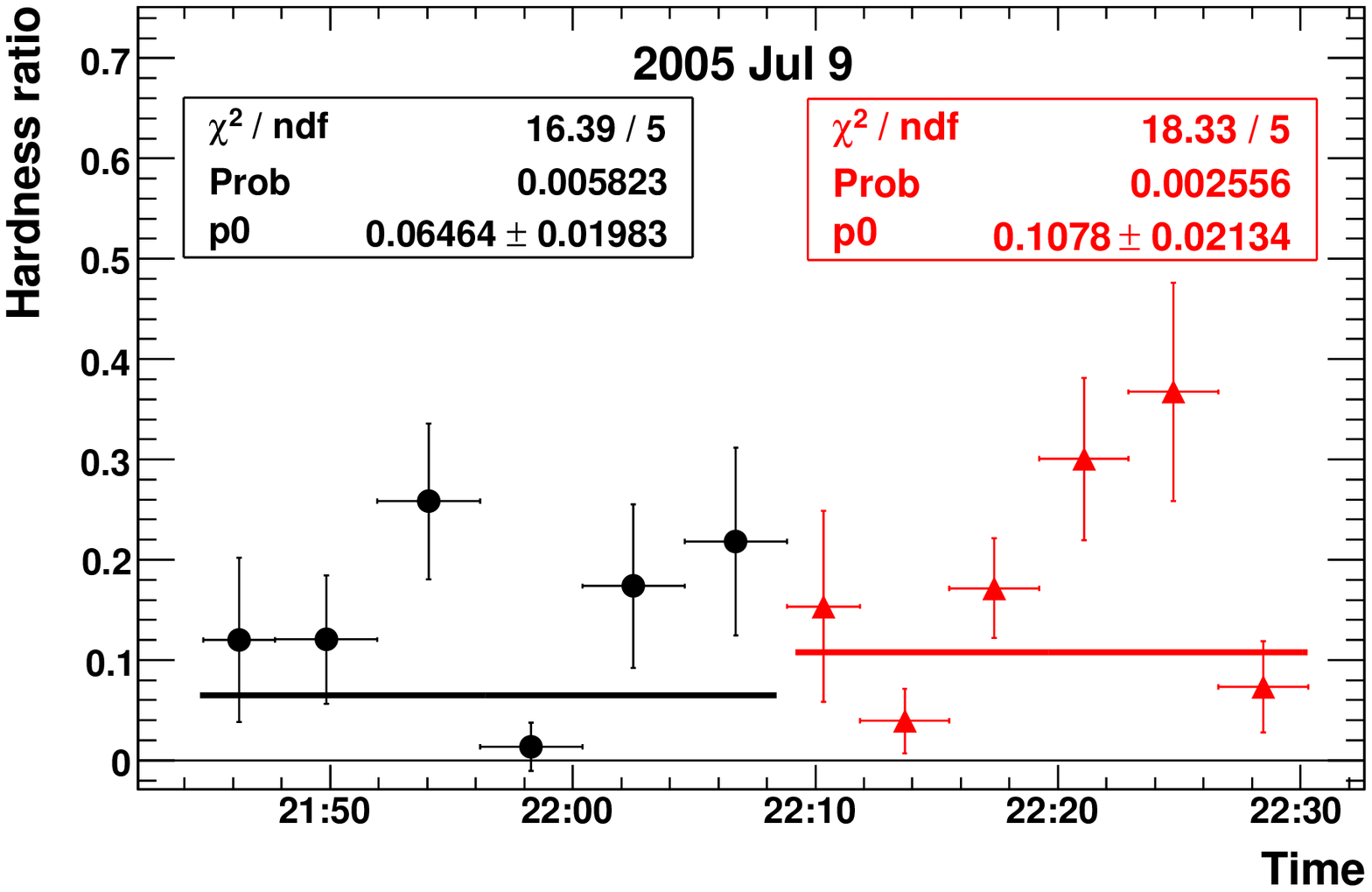}
\caption{Hardness ratio $F(1.2-10~ \TeV)/F(0.25-1.2~ \TeV)$ {\it vs} $Time$ for the nights of June 30 and July 9.
Horizontal bars represent the 4-minute time bins, and vertical bars denote 1$\sigma$ 
statistical uncertainties. 
Black squares and  red triangles denote pre-burst ('stable') and in-burst ('variable') 
emission respectively (see sect. \ref{Intradayvariations} and Fig. \ref{LCSingle}).   
The lines result from a constant fit to the data points (see insets for fit parameters).
\label{fig_HR_Time}}
\end{figure}

We also studied the time-evolution of the {\it hardness ratio}, defined as the ratio \\
\mbox{F(1.2-10 \TeV)/F(0.25-1.2 \TeV)} and which is computed 
directly from the LCs shown in Fig. \ref{LCSingle_MultiEnergyRanges_F0701} and 
Fig. \ref{LCSingle_MultiEnergyRanges_F0709}. The resulting 
graph is shown in Fig. \ref{fig_HR_Time}. The hardness ratios for the pre-burst and in-burst part are quantified 
by means of a constant fit. In both nights the hardness ratio is somewhat larger (1-2$\sigma$) in the in-burst 
than in the pre-burst part, in agreement with the observed spectral hardening (see Fig. 
\ref{fig_Intra-day-SpectralVariations}). It is worth noting that the hardness ratios for the pre-burst and 
in-burst time windows of June 30 are statistically compatible with being constant, those of July 9 
are much less so, as shown in the insets of Fig. \ref{fig_HR_Time}. 

\begin{figure}
\epsscale{1}
\plottwo{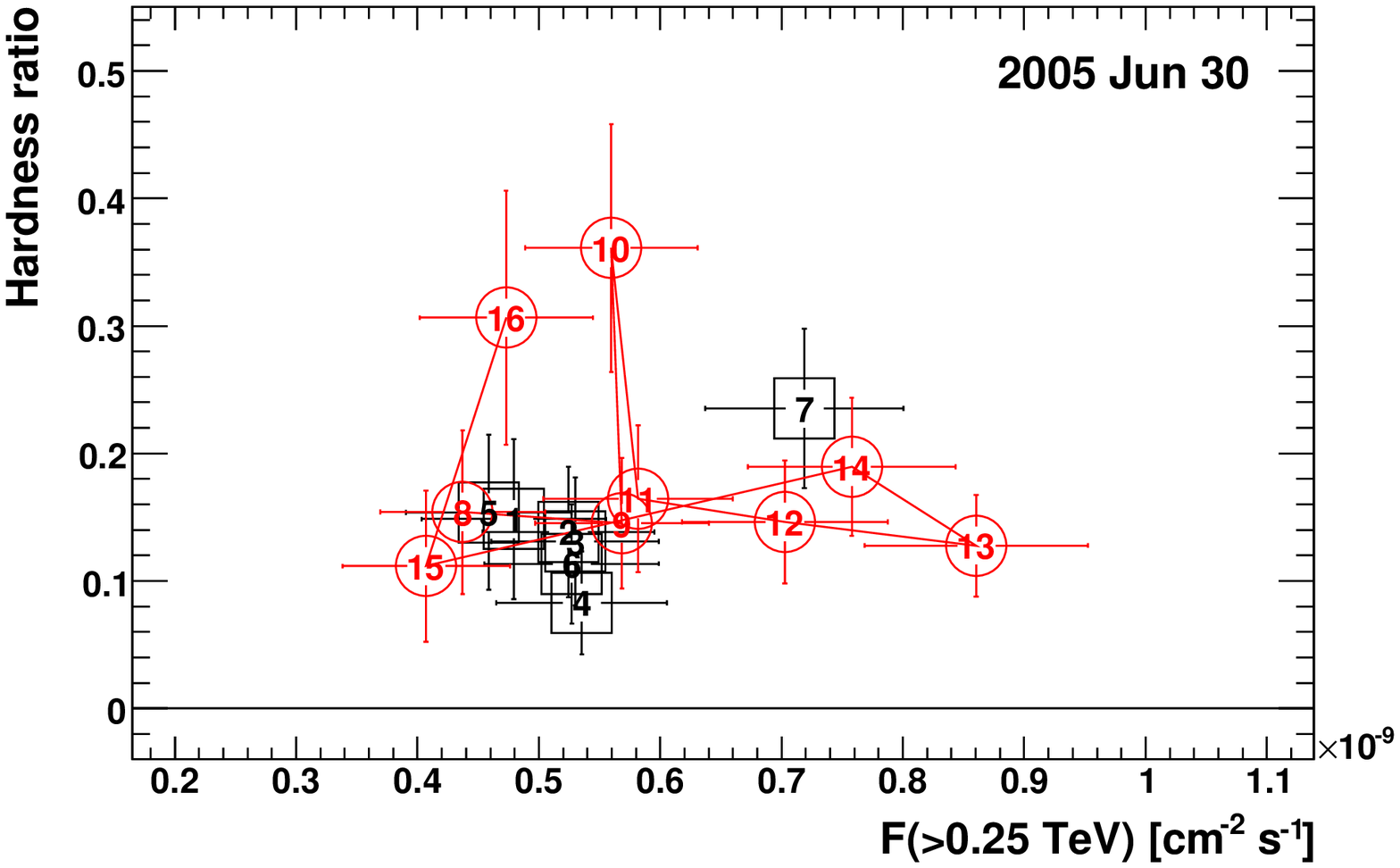}
{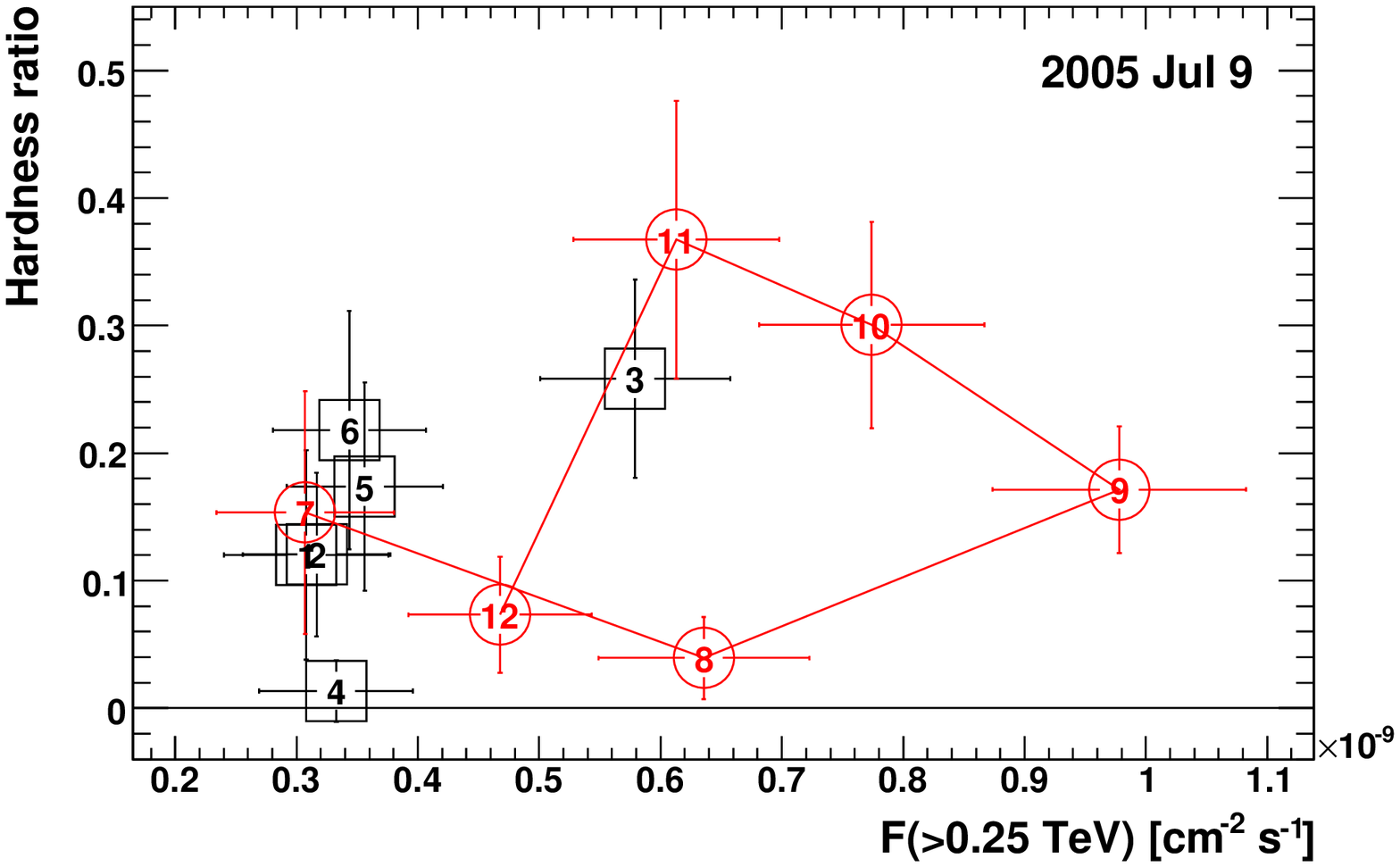}
\caption{Hardness ratio $F(1.2-10~ \TeV)/F(0.25-1.2~ \TeV)$ {\it vs} $F(>0.25~ \TeV)$ 
for the nights of June 30 and July 9.
Horizontal and vertical bars denote 1$\sigma$ 
statistical uncertainties. 
Black open squares and  red open circles denote pre-burst ('stable') and in-burst ('variable') 
emission respectively (see sect. \ref{Intradayvariations} and Fig. \ref{LCSingle}).   
The numbers inside the markers denote the position of the points in the LCs. 
The consecutive (in time) points of the in-burst LC are connected by red lines 
for better clarity.
\label{fig_HR_Flux}}
\end{figure}

The evolution of the hardness ratio with the emitted flux above \mbox{0.25 \TeV} is shown in 
Fig. \ref{fig_HR_Flux}. 
Both nights show some evidence for a larger spread in the in-burst part than in the pre-burst part. The evolution of 
the in-burst points from June 30, however, is somewhat chaotic, while the evolution of the in-burst points 
from July 9 shows a clear loop pattern rotating counterclockwise. The physical interpretation of this 
feature is given in section \ref{DiscSpectralShapeVariation}.
Concluding, a spectral hardening with 
increased emission characterizes the \vhe\ emission of \mbox{Mrk 501} also at short timescales.

\section{Discussion}
\label{Discussion}

In this section we discuss the \vhe\ light curve of Mrk 501 for May - July 2005
in comparison to
previous \iact\ observations made from the years 1997 through 2000. We also discuss the observed rapid 
flux variability in the framework of a basic model of gradual electron acceleration.
The broadband spectral features of Mrk 501 and the
intrinsic VHE spectra from different activity states are discussed
in the framework of a basic SSC model.
We also model the ensuing intrinsic \vhe\ spectra 
-- corresponding to different activity states -- in an SSC framework.

\subsection{Historical light curve}
\label{HistoricalLC}

\begin{figure}
\epsscale{1}
\plotone{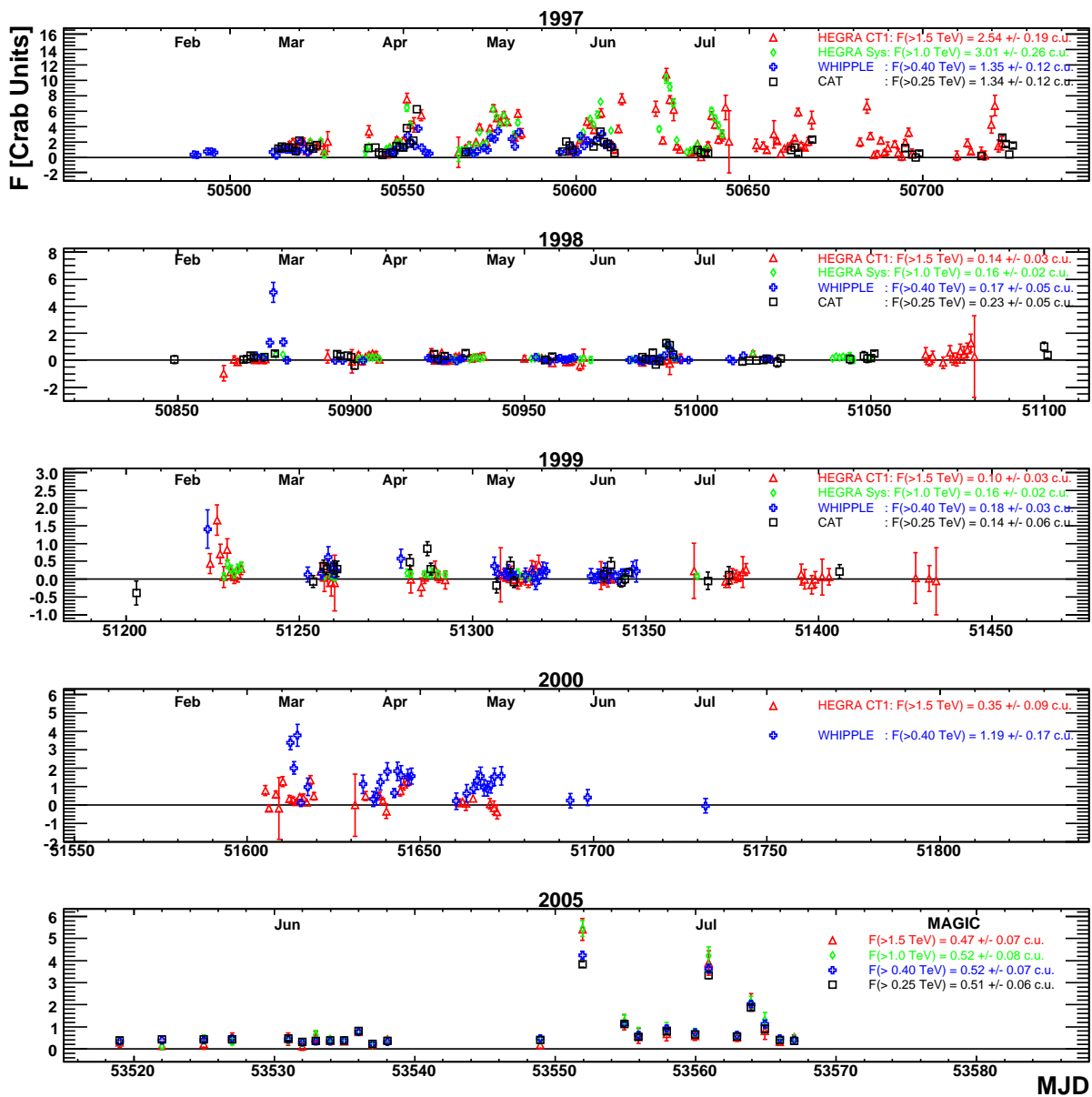}
\caption{Single-night \vhe\ LCs of \mbox{Mrk 501} obtained with various \iacts\ during several years. 
Vertical error bars denote 1$\sigma$ statistical uncertainties. Instruments and corresponding mean 
fluxes are reported for each observational campaign separately. The \magic\ data were reprocessed to 
match the energy ranges covered by previous instruments. 
\label{FigHistoricalLightCurve}}
\end{figure}

It is interesting to examine \mbox{Mrk 501}'s \vhe\ activity in 2005, as measured by \magic\, in 
perspective to that recorded in 1997-2000 by other IACTs such as HEGRA CT1 \citep{KranichPhD}, HEGRA 
CT System \citep{MartinWebPage}, Whipple \citep{Quinn1999}, and CAT \citep{Mrk501CAT,
Mrk501CATPironMoriond}. The long-term Mrk 501 lightcurve covering the years 1997 to 2005 is
shown in Fig. \ref{FigHistoricalLightCurve}. For easier comparison among instruments 
covering different energy ranges, the integral
flux values are given in Crab Units (c.u.). The mean fluxes for each year and 
instrument are shown in the insets
of Fig. \ref{FigHistoricalLightCurve}. They were derived using the exposure times as
statistical weights (whenever this information was available), 
and excluding data points $>$ 3$\sigma$ away from
each mean value, to permit a better comparison\footnote{
  Some HEGRA CT1, Whipple and CAT fluxes are negative, 
  unsurprisingly because many of the fluxes from these 
  instruments are just $\mincir$2$\sigma$ measurements. 
  For the HEGRA CT System we have no information about negative 
  values, hence the corresponding mean fluxes may be 
  slightly overestimated.}.
For MAGIC, mean fluxes were computed in the other IACTs' energy bands. In 1997, when 
\mbox{Mrk 501} was much brighter than the Crab, Whipple and CAT fluxes were in mutual agreement, but 
they significantly differed from the HEGRA CT System and HEGRA CT1 fluxes; 
the reasons being {\it (i)} the 
highest-flux nights (May--July 1997) happened to be covered by HEGRA CT and CT1 but not by Whipple and 
CAT; and {\it (ii)} the highest threshold energies of HEGRA CT and CT1, together with the 
fact that when \mbox{Mrk 501} is active  
its spectrum becomes significantly harder than the Crab's
(see Sects. \ref{SedCorrelation} and \ref{Sed4}), hence fluxes measured in Crab 
units are larger at higher energies. In 1998 and 1999 the various IACT 
data agreed on a flux of $\sim$0.15-0.20 \cu, i.e. an order of magnitude lower than in 1997. In 2000, 
March through May, the mean \vhe\ flux of \mbox{Mrk 501} increased to a level of 0.35$\pm$0.09 (HEGRA 
CT1) and 1.19$\pm$0.17 (Whipple), the main reason for the discrepancy being that HEGRA CT1 missed the 
nights with highest flux measured by Whipple. In 2005, May through July, the mean baseline \vhe\ flux 
from \mbox{Mrk 501} was $\sim$0.5 \cu, significantly lower than in 1997 and 2000, but higher than in 
1998 and 1999.

A 23-day flux periodicity was claimed by \citet{KranichPhD} using data from HEGRA CT1. 
Recently, \citet{Osone2006} 
confirmed the 23-day periodicity at \vhe\ frequencies and extended it to X-rays 
(based on ASM/RXTE data) using a more sophisticated time analysis.
This periodic 
modulation in the \vhe\ emission of \mbox{Mrk 501} may be evidence of a binary black
hole system with separation of the order of the gravitational radius, as 
suggested by \citet{Rieger2000,Rieger2003}, or selective absorption 
of \grays\ in the radiation of a hot spot orbiting in the inner 
part of the accretion disk, as pointed out by \citet{Bednarek1997b}.
This periodicity was not seen in the 1998-2000 campaigns, when the source 
was apparently not very active. This might indicate that such a periodicity 
in the emitted flux occurs only when the source shows very high activity. 
The MAGIC 2005 data did not have the required coverage for such a timing analysis.

\subsection{Interpretation of the measured fast flux variations and energy dependent time delays}
\label{DiscFastVariations}

The very short flux-doubling time and the energy-dependent time delays
of the flux variations can give us information on the acceleration
processes occurring in Mk501. In this section we argue that gradual
electron acceleration in the emitting plasma can provide a natural
explanation of the observed time structures.
\smallskip

\noindent
{\it i) Flare decay timescale.} 

Let us assume that the maximum energies of electrons accelerated in the relativistic
blob are determined by their radiation energy losses on synchrotron and IC processes, as
expected in the SSC model. The acceleration time should then equal the energy loss time
scale
\begin{eqnarray}
\tau_{\rm acc} ~=~ \tau_{\rm cool},
\label{eq1}
\end{eqnarray}
\noindent
where
\begin{eqnarray}
\tau_{\rm acc} ~=~ E_{\rm e}'/{\dot P}_{\rm acc} = E_{\rm e}'/(\xi c E_{\rm e}'/R_{\rm L})
\approx 0.1\,E_{\rm e}'/\xi B~~~{\rm s}.
\label{eq1b}
\end{eqnarray} 
\noindent
Here $E_{\rm e}'$ is the electron energy (in \TeV) in the blob, ${\dot P}_{\rm acc}$ is the 
rate of energy gain during the acceleration process, $B$ (in G) is the magnetic field strength 
in the acceleration region, $R_{\rm L}$ is the electron Larmor radius, and $\xi$ is the 
acceleration efficiency. The cooling time of electrons can be expressed by 

\begin{eqnarray}
{{1}\over{\tau_{\rm cool}}} = {{1}\over{\tau_{\rm syn}}} + {{1}\over{\tau_{\rm IC}}}.
\label{eq2}
\end{eqnarray}
\noindent
Then, the cooling time can be expressed by only the synchrotron cooling time
\begin{eqnarray}
\tau_{\rm cool} = \tau_{\rm syn}/(1 + \eta),
\label{eq2a}
\end{eqnarray}
\noindent
where  $\eta \equiv \tau_{\rm syn}/\tau_{\rm IC}$, 
$\tau_{\rm syn} ~=~ E_{\rm e}'/{\dot P}_{\rm syn}$,
${\dot P}_{\rm syn} = 4\pi\sigma_{\rm T}cU_{\rm B}\gamma_{\rm e}^2/3$,
$\sigma_{\rm T}$ is the Thomson cross section, $c$ the velocity of light, 
$U_{\rm B} = B^2/8\pi$ is the energy density of the magnetic field, $\gamma_{\rm e} 
= E_{\rm e}'/mc^2$, and $m$ is the electron rest mass. The parameter $\eta$ 
corresponds to the ratio of the power emitted by electrons in IC 
and synchrotron processes, respectively (which is reciprocal to the coresponding
cooling times $\tau_{\rm syn}/\tau_{\rm IC}$). The \magic\ \gray\ and 
corresponding RXTE/ASM \xray\ data permit to constrain $\eta$  to 
$\lesssim$ 0.7.
The modelling of the SED presented in section \ref{EBLCorr} 
(see, Fig. \ref{fig_OverallSED_EBL_Corrected}), however, suggests that $\eta$ is more likely of 
the order of $\sim 0.2$. We therefore used $\eta$=0.2 in all the following estimates.
By comparing Eqs.~\ref{eq1},\ref{eq1b} and~\ref{eq2a} and setting $E_{\rm e}'\approx E_{\rm 
TeV}/\delta$ \TeV\ (where $\delta$ is the Doppler factor of the relativistically-moving 
emitting plasma blob and $E_{\rm TeV}$ is the electron energy (in \TeV) in the 
observer's frame, we obtain the condition on the acceleration 
efficiency of electrons,
\begin{eqnarray}
\xi ~\approx~ 10^{-3}B E_{\rm TeV}^2(1+\eta) \delta^{-2}.
\label{eq3}
\end{eqnarray}
\noindent
If the observed decay timescale of the flare, $\tau_{\rm f}$, is also due 
to radiative processes, then,
\begin{eqnarray}
\tau_{\rm cool} ~=~ \tau_{\rm f} \delta\,.
\label{eq3b}
\end{eqnarray}
\noindent
Inserting Eq. \ref{eq2a} into Eq. \ref{eq3b}, and expanding 
$\tau_{\rm syn}$ and ${\dot P}_{\rm syn}$, permits to 
estimate $B$ in the cooling region: 
\begin{eqnarray}
B ~\approx~ 11.2[(1 + \eta) \cdot (E_{\rm TeV}\tau_{\rm f})]^{-1/2}~~~{\rm G},
\label{eq3c}
\end{eqnarray}
\noindent
with $\tau_{\rm f}$ in seconds. If electron acceleration and cooling are 
co-spatial, Eq.~\ref{eq3} becomes
\begin{eqnarray}
\xi ~\approx~ 1.1\times 10^{-2} {{E_{\rm TeV}^{3/2} \cdot (1+\eta)^{1/2}}\over{\tau_{\rm f}^{1/2}\delta^2}}
\label{eq4}
\end{eqnarray}
\noindent
For the parameters of the July 9 flare Mrk501, the {\it characteristic flux variability time}
$\tau_{\rm f} \approx 3$ minutes (see Sect. \ref{Intradayvariations}) 
and $E_{\rm TeV} \approx 1$, we obtain $B\approx 0.8$ G 
and $\xi\approx 0.9\times 10^{-3} \delta^{-2}$. Note, that this estimate of the magnetic field in the 
emission (acceleration) region of Mrk 501 is consistent with previous estimates, 
based on a homogeneous SSC model and the assumption of a very short variability 
timescale such as for the April 15-16 1997 flare (see e.g., Fig 3c in \citet{Bednarek1999}). 
\smallskip

\noindent
{\it ii) Energy-dependent time delay in peak flare emission.} 

The time delay between the peaks of F($<$0.25 \TeV) and F($>$1.2 \TeV) during the 
July 9 flare, can be interpreted as due to the gradual acceleration of electrons in the relativistic 
blob. As reported in section \ref{Intradayvariations}, 
under the assumption that the shape of the flares is the same in the 
two energy ranges, the time delay is $\Delta\tau_{\rm H-L} = 4 \pm 1$ minutes.

Within the above framework, the time delay should correspond to the difference 
between the acceleration times of electrons to energies $E_{\rm TeV}^{\rm L}\approx 
0.25$ and $E_{\rm TeV}^{\rm H}\approx 1.2$, according to
\begin{eqnarray} 
\tau_{\rm acc}^{\rm H} - \tau_{\rm acc}^{\rm L} ~=~ \Delta\tau_{\rm H-L} \delta.
\label{eq5}
\end{eqnarray}
\noindent 
Assuming for simplicity that the electron energies are determined by the energies of 
the emitted \vhe\ radiation, i.e. ${E_{\rm e}^{'{\rm L}}}$$\approx$$E_{\rm TeV}^{'{\rm L}}/\delta$ and 
${E_{\rm e}^{'{\rm H}}}$$\approx$$E_{\rm TeV}^{'{\rm H}}/\delta$, we can use Eq.~\ref{eq1b} 
to model the acceleration time of electrons in eq.\ref{eq5}. Finally, by reversing 
Eq.~\ref{eq5}, we get another limit on the electron acceleration efficiency,
\begin{eqnarray}
\xi ~\approx~ 0.1{{(E_{\rm TeV}^{\rm H} - E_{\rm TeV}^{\rm L})}\over{B \delta^2 \Delta\tau_{\rm H-L}}}.
\label{eq6}
\end{eqnarray}
Applying the observed values of $E_{\rm TeV}^{\rm H}$, $E_{\rm TeV}^{\rm L}$, 
$\Delta\tau_{\rm H-L}$ and the estimate of $B$ from Eq.~\ref{eq3c} 
(which is valid if electron acceleration and cooling are co-spatial) we obtain 
$\xi$$\approx$$0.5\times 10^{-3}\delta^{-2}$, which is roughly the same  
value as determined above using the exponential flux decay time. 
Note that diffusion and/or spatial dishomogeneities might increase the 
volume where electrons cool down, thus making the magnetic field 
in the cooling region lower than that of the acceleration region.
That would imply that the estimated values for the parameter $\xi$
using equations \ref{eq4} and \ref{eq6} are upper and lower limits, respectively.

We conclude that the time delay of the flare peak emission in different ranges of energy 
can result from the gradual acceleration of the emitting electrons 
in the blob. The inferred blazar Doppler factors, $\delta\sim 10-15$ 
(e.g. \cite{Costamante2002}), imply a relatively 
inefficient acceleration, $\xi\sim 10^{-5}$. This value is significantly lower than 
required by the observations of $\gamma$-ray emission from the pulsar wind nebulae 
in which leptons are accelerated in the shock wave in the relativistic pulsar wind. 
For example, in the Crab Nebula $\xi$ has to be of the order of $\sim 0.1$, since 
leptons are accelerated clearly above $10^3$ \TeV\ (approximately $10\%$ of the maximum 
available potential drop through the pulsar magnetosphere). Therefore, the physics 
of the acceleration process in the relativistic jets of BL Lacs and  relativistic 
shocks in the pulsar wind nebulae may differ significantly. 

A somewhat more speculative issue that blazar emission permits to explore
concerns non-conventional physics. 
Energy-dependent arrival times are predicted by several models of Quantum Gravity, 
which quantify the first-order effects of the violation of Lorentz symmetry. One could, 
therefore, speculate that the observed time difference is explained by such models, 
although source-inherent effects could certainly not be excluded. A more detailed 
investigation of such interpretations of our data is still going on.

\subsection{Interpretation of the spectral shape variations}
\label{DiscSpectralShapeVariation}

The observed correlation between spectral shape and (bolometric) luminosity is naturally 
accounted for in the SSC scenario. \citet{Pian1998,Tavecchio2001} discuss the SED 
variations of \mbox{Mrk 501} during the giant 1997 flare. During the 1997 flare the 0.1-200 keV 
band synchrotron spectrum became exceptionally flat (photon spectral index $a$$<$1), 
peaking at $\magcir$100 keV -- a shift to higher frequencies by a factor of $~$100 from 
previous, more quiescent states. The \vhe\ data (from the Whipple, HEGRA, and CAT telescopes) 
showed a progressive hardening from the baseline state ($a$$>$2) through a more active state
to a flaring state ($a$$\sim$2). In the SSC scenario, these flux-dependent spectral changes 
implied that a drastic change in the electron spectrum caused the increase in emitted power: 
a freshly injected electron population has a flatter high-energy slope and a higher maximum 
energy than an aging population, which cause a shift of the SED to higher frequencies. 

In section \ref{Sedintradayvariations} we reported that the spectrum of \mbox{Mrk 501} not 
only hardens on long time scales, with the overall emitted flux, but also during the 
shorter events of June 30 and July 9. The burst from July 9 showed a remarkable 
variability, and the evolution of the hardness ratio with the flux 
(right-hand plot of Fig \ref{fig_HR_Flux}) contains valuable information 
about the dynamics of the source. In the pre-burst phase, the hardness ratio does not vary 
significantly; yet during the burst phase, it varies following a clear loop pattern rotating 
counterclockwise. As pointed out by \citet{kirk1999},  one expects to have 
this behaviour for a flare where the variability, acceleration and cooling timescales are similar;
which implies that, during this flare, the dynamics of the system is dominated by 
the acceleration processes, rather than by the cooling processes. 
Consequently, the emission propagates from lower to higher energy, so the lower 
energy photons lead the higher energy ones (that is the so-called hard lag).
This indeed agrees well 
with the argumentation given in 
section \ref{DiscFastVariations}, where the time delay between $E_{>1.2 ~TeV}$ and $E_{<0.25 ~TeV}$ 
is shown to be consistent with the gradual acceleration of the electrons. 


In a systematic 
study performed by \citet{gliozzi2006} using \xray\ data from 1998 to 2004, 
this behaviour was not observed on more typical (longer) flares, where actually the opposite 
behaviour (clockwise rotation) was indicated. This might point to the fact that 
these physical processes might be responsible only for the shortest flux variations, 
and not for the variability on longer timescales.

\subsection{Interpretation of the increased variability with energy}
\label{DiscVariability}

In the SSC framework, the variability observed in the \vhe\ emission brings information 
about the dynamics of the underlying population of relativistic electrons 
(and possibly positrons). In this context, the general variability trend reported 
in section \ref{Quantificationvariability} is 
interpreted by the fact that the \vhe\ \grays\ (as well as the \xrays) are produced 
by more energetic particles, that are characterized by shorter cooling timescales;
causing the higher variability amplitude observed at the highest energies.
It is worth noticing that such an injection of high energy particles would produce 
a shift in the IC peak, which is indeed observed during this observing campaign, 
as reported in sections \ref{Sed} and \ref{EBLCorr}.

\subsection{Spectra corrected for the  EBL absorption }
\label{EBLCorr}

\begin{figure}
\epsscale{1}
\plottwo{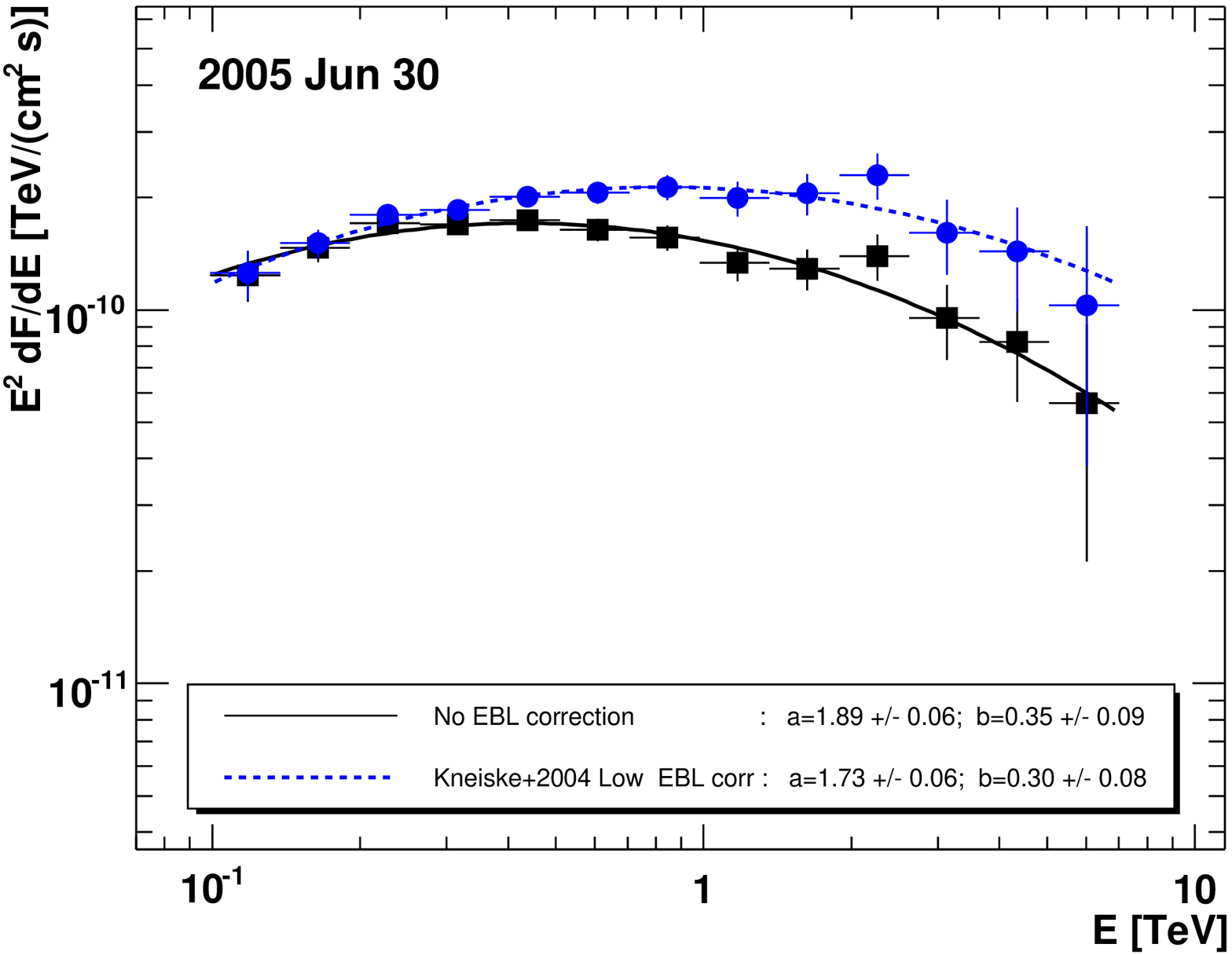}
{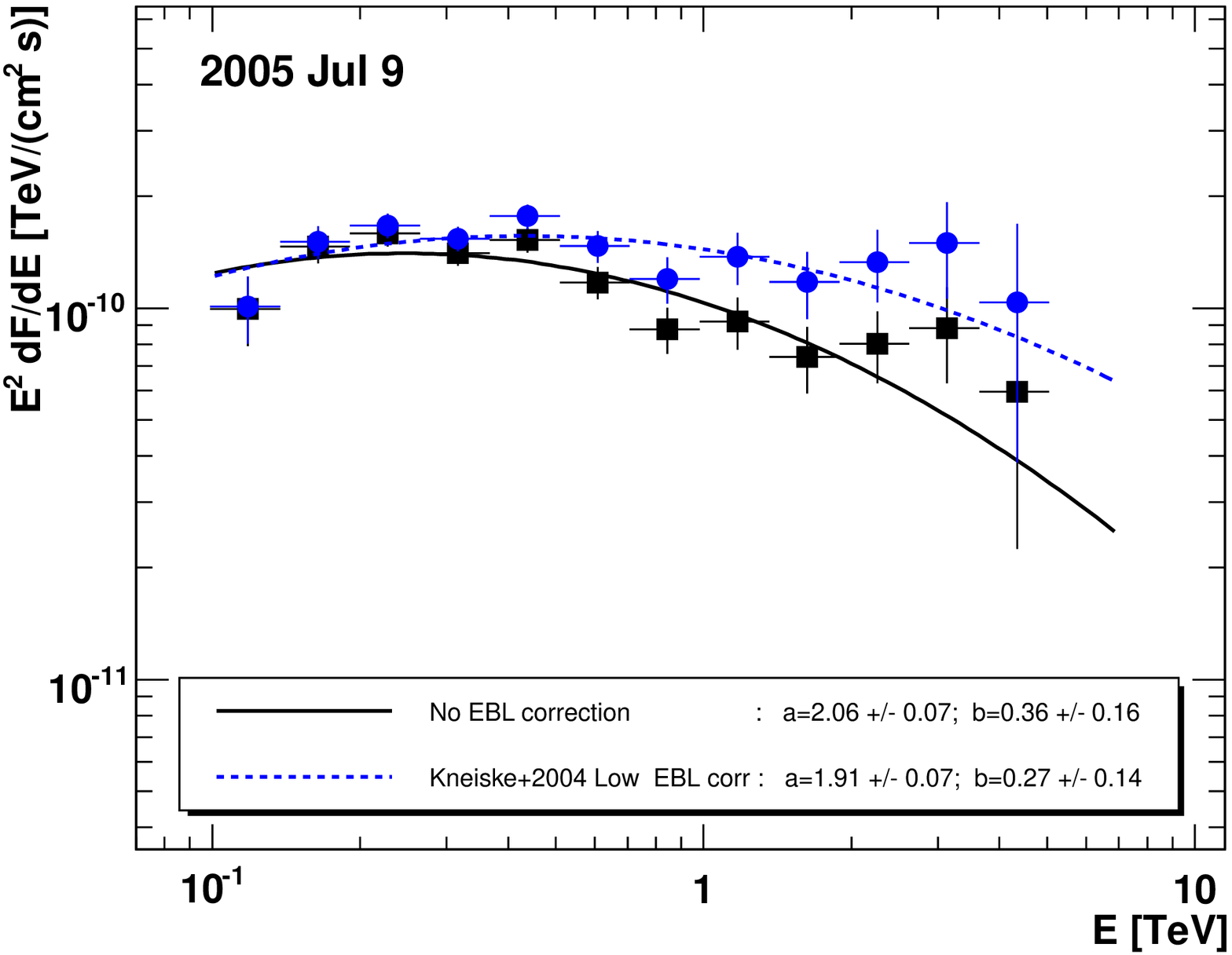}
\caption{
Spectra of \mbox{Mrk 501} in the nights of June 30 (left) and July 9 (right) when the source was flaring. Bars as in 
Fig. \ref{Sed4groups}. The spectra have been corrected for EBL absorption using \citep{EBLKneiske}'s 'Low' EBL model.
The curves show log-parabolic fits (see eq. \ref{eq_powerlawEDepInd}) whose corresponding 
parameters are reported in the insets.
\label{fig_flarenights_EBL_Corrected}}
\end{figure}

In this section we correct the measured spectra for the EBL absorption. To this purpose we use 
Kneiske et al. (2004: 'Low'). A correction using the EBL models Aharonian et al. (2006) and 
and Primack et al. (2005) gave very similar results, while using Kneiske et al. (2004: 'Best') 
and Stecker et al. (2006,2007) gave slightly larger energy fluxes above  1 \TeV\footnote
{The difference increases with the energy, being $\sim$50\% at 5 \TeV.}.
Fig. \ref{fig_flarenights_EBL_Corrected} shows the spectra from the active
nights June 30 and July 9 before/after correction for the EBL absorption. 
In spite of our proximity to Mrk501, the effect of the EBL is not negligible, 
and the spectral peak moves to higher energies.

\begin{figure}
\epsscale{0.5}
\plotone{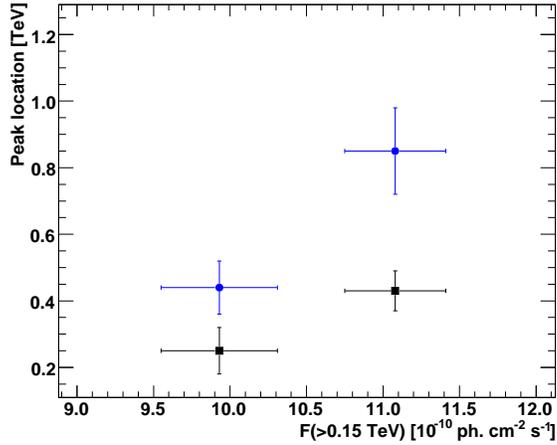}
\caption{Spectral peak location versus flux above 0.15 \TeV\ for the two flaring nights 
(June 30 and July 9).  The spectra were fitted with eq. 
\ref{eq_powerlawEDepInd} (see Fig. \ref{fig_flarenights_EBL_Corrected}) 
and the peak location and its associated error were calculated using 
eqs. \ref{eq_SED_Peak} and \ref{eq_SED_PeakError}. 
The black squares correspond to the observed spectra and 
the blue circles correspond to the spectra after correction for 
the EBL absorption using \citep{EBLKneiske}'s 'Low' EBL model. 
\label{FigCorrPeakPosGammaFlux}}
\end{figure}

The location of the spectral peaks (calculated using eqs. \ref{eq_SED_Peak} and \ref{eq_SED_PeakError}) 
are shown  as a function of F ($>$0.15 \TeV) in Fig. \ref{FigCorrPeakPosGammaFlux}
for the flaring nights before and after the EBL correction. The figure seems to indicate 
a displacement of the peak location with the increasing flux, yet the error bars 
are too large to be conclusive.  
On the other hand the peak location is certainly at $<$0.1 \TeV\ when Mrk 501 is 
in a low state (see sect. \ref{Sed4}). 
Hence, there is evidence for an overall peak location versus luminosity trend.

\begin{figure}
\epsscale{1}
\plotone{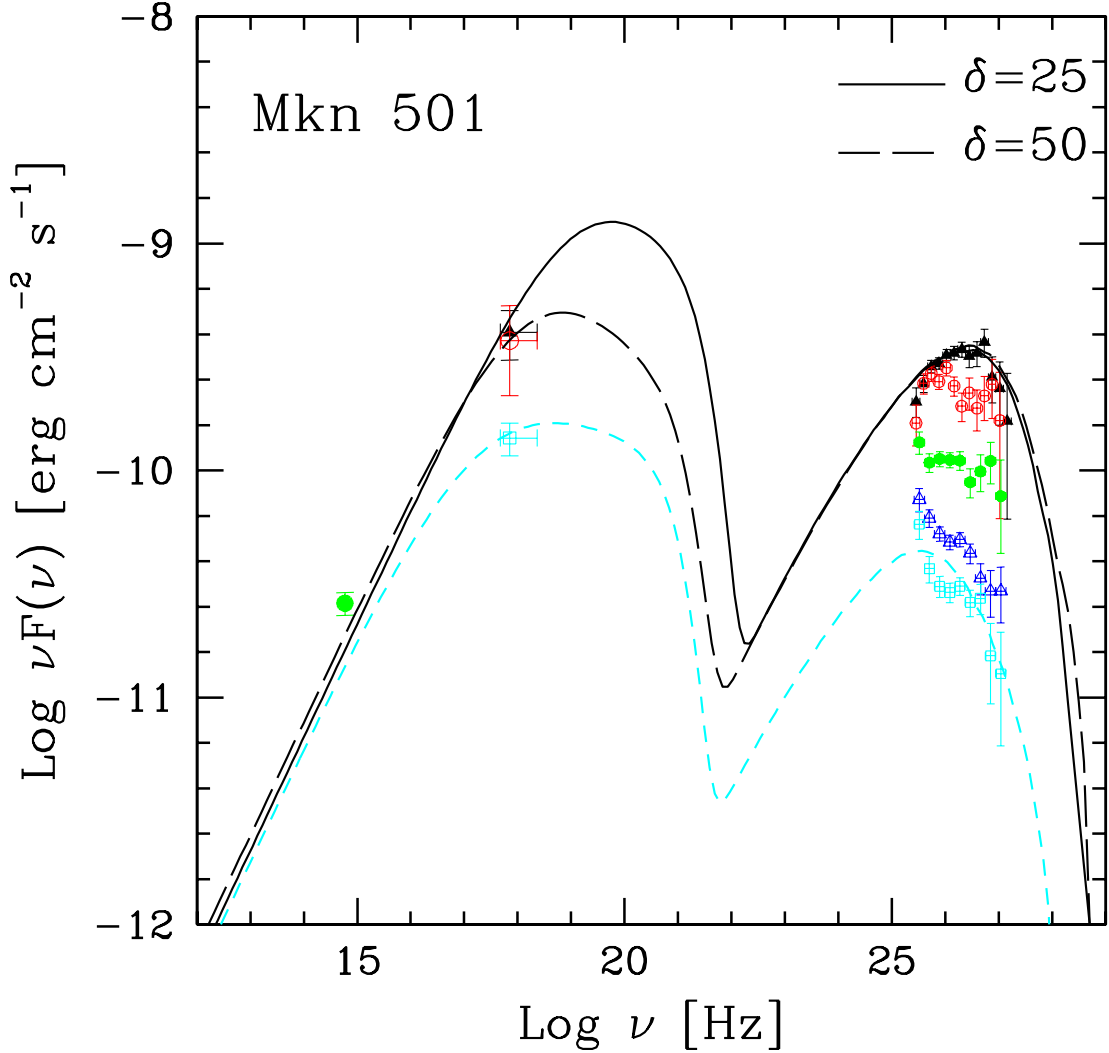}
\caption{Overall SED from Mrk501. The optical data from the  KVA Telescope is represented 
with a green full circle; the \xray\ data from \asm\ depicted with a black full triangle for June 30, 
red open circle for July 9, and light blue open square for the other nights (combined); 
the \vhe\ data from \magic\ are represented as 
black full triangles (June 30), red open circles (July 9), green full circles ('high flux' data-set), 
dark blue  open triangles ('medium flux' data-set), and light blue open squares ('low flux' data-set). See 
Sect. \ref{Sed4} for definitions of high, medium and low flux data-sets. Vertical error bars 
denote 1$\sigma$ statistical uncertainties. The \vhe\ spectra are corrected for EBL extinction 
using \citep{EBLKneiske}'s 'Low' EBL model. The highest and the lowest state were fit with a 
one zone SSC model (described in \citet{Tavecchio2001}). See table \ref{SSCModelParams} 
and section \ref{EBLCorr} for details of the modeling.
\label{fig_OverallSED_EBL_Corrected}}
\end{figure}

Fig. \ref{fig_OverallSED_EBL_Corrected} shows the June 30 and 
July 9 spectra, as well as the mean 'high', 'medium', and 'low'-flux spectra (see sect. \ref{Sed4}), 
EBL-de-absorbed using the 'Low' model of \citet{EBLKneiske}. The \xray\ fluxes measured by \asm, 
and the optical flux observed by the KVA Telescope, are also shown. The optical flux 
from the host galaxy, estimated by \citet{Nilsson2007} to be 12.0 $\pm$ 1.2 mJy, has been 
subtracted.

The best-fitting one-zone, homogeneous SSC models of Mrk501's intrinsic spectrum for the highest 
state of the source (corresponding to the active night of June 30) and for the lowest state (the 
mean spectrum corresponding to the 'low'-flux bin, see Fig. \ref{Sed4groups}) are displayed in Fig. 
\ref{fig_OverallSED_EBL_Corrected}. The fit parameters (electron population's break and max/min 
energies, high/low-$E$ spectral slopes, normalization: $E_{\rm br}$, $E_{\rm max}$, $E_{\rm min}$, 
$n_1$, $n_2$, $K$; plasma blob's radius, magnetic field, Doppler factor $R$, $B$, $\delta$) are 
reported in Table \ref{SSCModelParams} (see Tavecchio et al. 2001 for details on the model). It 
should be noted that two different fits to the high-state spectrum are possible -- whose main 
differing parameters are, respectively, $\delta$,$B$=$25$, $0.23~G$ 
(solid black curve) and $50$, $0.053~G$  (dashed black curve) -- which show that fairly different synchrotron peaks are 
possible within our X-ray and (EBL-corrected) \TeV\ data. Spectrally more extended X-ray data would 
probably have solved the degeneracy. The optical data, too, do not lift the degeneracy: once the 
optical light contribution of the underlying host galaxy is subtracted, the observed energy flux 
is rather compatible with the SED models for the different 
activity states\footnote{
The measured optical flux might have contributions from regions outside the one producing the radiation at 
X-ray and \gray\ frequencies.}.
The fit to the 'low'-state spectrum is characterized, perhaps unsurprisingly, by 
a change of the internal physical conditions of the emitting plasma blob rather than by a change of 
its bulk attributes (blob size and relativistic Doppler factor): the low state is characterized, 
with respect to the high state, by lower max/break energies and normalization of the electron 
population and by a somewhat stronger magnetic field. One nice consistency feature of all the fits is that, 
in all cases, the radius of the plasma blob, $R$$=$$10^{15}$ cm, implies a crossing time $t_{\rm 
cr}$$=$$R/\delta c$, comparable to that inferred from the observed duration ($\sim$20 minutes) of 
the flare, $\Delta$$t_{\rm flare}$. 

The SED models for Mrk 501 derived and discussed in this section 
can be compared with some previous published models, like Pian et al. (1998): $\delta,B$$
\simeq$$15,0.8$G from one-zone SSC modeling of 1997 SEDs, with flatter/steeper electron 
distributions for active/quiescent phases; Tavecchio et al. (1998): $\delta,B$$\simeq$$8-20,
0.5-1$G from one-zone SSC modeling of historical quiescent SED and $\delta,B$$\simeq$$7,1$G 
(and very high break energy) for the active SED; Bednarek \& Protheroe (1999): $\delta,B$$
\simeq$$12-36, 0.07-0.6$G from 1997 SEDs modeled with one-zone SSC 
requiring \gray\  transparency of the emitting blob; Kataoka et al. (1999): 
$\delta,B$$\simeq$$15,0.2$G from one-zone SSC modeling of simultaneous 1996 SED; Katarzy\'nski 
et al. (2001): $\delta,B=14,0.2$G from SSC modeling of non-simultaneous broadband SED. 
The SED of Mrk 501 can also be modelled by less conventional approaches, requiring 
magnetic fields in the emission region smaller than $0.005$G, \citet{Krawczynski2007}.
We should, however, remember that for this highly variable source, constraints derived for some epochs 
may not apply to other epochs (the simple one-zone model has 9 free parameters!): in particular, 
most published models refer to the giant 1997 flare hence comparisons with our results may not 
be straightforward. However, Tavecchio et al. (2001) modelled different emission states of 
Mrk501 in 1997, 1998, and 1999 by just changing the electron energy distribution (slopes, 
break energy, number density) and keeping the other parameters frozen, similarly to what was 
done here (see in Table \ref{SSCModelParams} the two states represented by solid lines in Fig. 
\ref{fig_OverallSED_EBL_Corrected}). It is also worth mentioning the work done by 
\citet{Krawczynski2002}, in which the energy spectra (for different days) were modelled using a 
time-dependent code. 

It should be remarked that in TeV blazars, while the bright and rapidly variable VHE emission 
implies that at the scales where this emission originates ($\sim$0.1 pc from the jet apex) 
the jet is highly relativistic ($\delta$$\sim$10-20 with no EBL-absorption correction, and 
$\delta$$\mincir$50 with correction), at VLBI ($\sim$1 pc) scales the jets are relatively slow 
(see Ghisellini et al. 2005, and references therein). Hence, to reconcile the high $\delta$'s 
derived from VHE data with the much lower $\delta$'s derived from VLBI radio measurements, 
the jets of TeV blazars must either undergo severe deceleration (Georganopoulos \& Kazanas 
2003) or be structured radially as a two-velocity, inner spine plus outer layer, flow 
(Ghisellini et al. 2005).

\section{Concluding remarks}
\label{Conclusion}

In this work, we have undertaken a systematic study of the temporal and spectral variability 
of the nearby blazar \mbox{Mrk 501} with the \magic\ telescope at energies $>$ 0.1 \TeV.
During 24 observing nights between May and July 2005, 
all of which yielded significant detections, we measured 
fluxes and spectra at levels of baseline activity ranging from $<$0.5 
to $>$1 \cu. During two nights, on June 30 and July 9, \mbox{Mrk 501} 
underwent into a clearly active state with a \gray\ emission $>$3 \cu, 
and flux-doubling times of 
$\sim$2 minutes. The $\sim$20-minute long flare of July 9 showed an indication of a 
4$\pm$1 min time delay between the peaks of F($<$0.25 TeV) and F($>$1.2 TeV), which
may indicate a progressive acceleration of electrons in the emitting plasma blob. 
An overall trend of harder 
spectra for higher flux is clearly seen on intra-night, night-by-night, and longer-term timescales.
The \vhe\ \gray\ variability was found to increase with energy, regardless whether 
the source is in active or quiescent state, and it is significantly higher 
than the variability at \xray\ frequencies. 
A spectral peak, at a location dependent on source luminosity, was clearly observed during the active 
states. 
All these features are 
naturally expected in synchro-self-Compton (SSC) models of blazar \vhe\ emission.
There are no simultaneous good quality X-ray measurements during the \magic\ observations. 
As a consequence, the SSC 
model of the X-ray/\vhe\ SED of \mbox{Mrk 501} in an active state is not unequivocally 
constrained, but it still restricts 
the emitting plasma blob to have Doppler factors in range of 25-50 
and magnetic fields in the range 0.05-0.5$\,$G.

\section{Acknowledgments}

We would like to thank the IAC for the excellent working conditions on the La Palma
Observatory Roque de los Muchachos. We are grateful to the ASM/RXTE team for their
quick-look results. The support of the German BMBF and MPG, the Italian INFN and the
Spanish CICYT is gratefully acknowledged. This work was also supported by ETH Research
Grant TH-34/04-3 and by Polish Grant MNiI 1P03D01028. Besides, the authors want to 
thank Deirdre Horan,  Martin Tluczykont and  Fr\'{e}d\'{e}ric Piron for providing 
Mrk501 data from WHIPPLE, HEGRA CT1, HEGRA CT System and CAT, respectively, 
and for useful discussions.


\begin{deluxetable}{cccccccccc}
\tabletypesize{\scriptsize}
\tablecaption{\magic\ observation of \mbox{Mrk 501} \label{SummaryTable}}
\tablehead{
\colhead{MJD} &\colhead{$T_{obs}$ \tablenotemark{a}}&\colhead{ZA\tablenotemark{b}}
&\colhead{$S_{comb}$\tablenotemark{c}} 
& \colhead{$F_{>0.15~TeV}$\tablenotemark{d}}
& \colhead{$F_{>0.15~TeV}$}
& \colhead{$K_0$\tablenotemark{e}}
& \colhead{$a$\tablenotemark{f}} 
& \colhead{${\chi^2}/{NDF}$\tablenotemark{g}} & \colhead{$P$}\tablenotemark{h} \\
\colhead{Start} & \colhead{($h$)} & \colhead{($deg$)} & \colhead{sigma}
& \colhead{($\frac{10^{-10} ~ph}{cm^2 \cdot s}$)} &\colhead{{\it (Crab Units)}}
& \colhead{($\frac{10^{-10} ~ ph}{cm^2 \cdot s \cdot 0.3 TeV}$)}
& \colhead{} & \colhead{} &\colhead{(\%)}}
\startdata

53518.980 & 0.75 & 19.10-28.95 & 6.44 & 1.19 $\pm$ 0.25 & 0.37 $\pm$ 0.08 & 2.63 $\pm$ 0.48 & 2.17 $\pm$ 0.25 & 2.7/8 &  95.2  \\
53521.966 & 1.85 & 9.97-30.10 & 8.90 & 1.51 $\pm$ 0.17 & 0.47 $\pm$ 0.05 & 2.94 $\pm$ 0.33 & 2.61 $\pm$ 0.16 & 10.8/7 &  15.0  \\
53524.969 & 0.58 & 19.18-27.73 & 6.98 & 2.04 $\pm$ 0.29 & 0.64 $\pm$ 0.09 & 3.71 $\pm$ 0.53 & 2.47 $\pm$ 0.23 & 1.6/6 &  95.0  \\
53526.975 & 0.98 & 9.96-28.94 & 8.69 & 1.63 $\pm$ 0.22 & 0.51 $\pm$ 0.07 & 3.26 $\pm$ 0.38 & 2.49 $\pm$ 0.17 & 3.8/9 &  92.4  \\
53530.973 & 0.47 & 15.22-22.32 & 6.52 & 1.53 $\pm$ 0.32 & 0.48 $\pm$ 0.10 & 2.28 $\pm$ 0.65 & 1.97 $\pm$ 0.49 & 1.1/3 &  78.9  \\
53531.959 & 0.90 & 15.21-25.15 & 6.98 & 1.29 $\pm$ 0.24 & 0.41 $\pm$ 0.07 & 2.69 $\pm$ 0.38 & 2.57 $\pm$ 0.30 & 9.1/6 &  16.6  \\
53532.936 & 0.53 & 23.80-30.11 & 5.44 & 1.50 $\pm$ 0.28 & 0.47 $\pm$ 0.09 & 2.41 $\pm$ 0.53 & 2.34 $\pm$ 0.36 & 1.2/7 &  99.2  \\
53533.933 & 1.63 & 12.85-30.09 & 7.83 & 1.44 $\pm$ 0.17 & 0.45 $\pm$ 0.05 & 2.46 $\pm$ 0.32 & 2.55 $\pm$ 0.19 & 10.3/8  & 24.2  \\
53534.940 & 2.07 & 9.95-30.09 & 9.56 & 1.43 $\pm$ 0.15 & 0.45 $\pm$ 0.05 & 2.71 $\pm$ 0.27 & 2.68 $\pm$ 0.16 & 8.9/9 &  44.8  \\
53535.934 & 3.43 & 9.95-30.07 & 18.58 & 2.69 $\pm$ 0.13 & 0.85 $\pm$ 0.04 & 4.45 $\pm$ 0.24 & 2.42 $\pm$ 0.06 & 11.9/12  & 45.3  \\
53536.947 & 2.68 & 9.95-29.93 & 7.01 & 0.75 $\pm$ 0.13 & 0.24 $\pm$ 0.04 & 1.36 $\pm$ 0.21 & 2.73 $\pm$ 0.29 & 5.7/7 &  57.1  \\
53537.971 & 3.08 & 9.95-30.10 & 11.52 & 1.25 $\pm$ 0.10 & 0.39 $\pm$ 0.03 & 2.08 $\pm$ 0.19 & 2.46 $\pm$ 0.14 & 8.2/8 &  41.4  \\
53548.931 & 0.87 & 9.98-20.68 & 6.12 & 1.21 $\pm$ 0.25 & 0.38 $\pm$ 0.08 & 2.39 $\pm$ 0.38 & 2.28 $\pm$ 0.27 & 0.6/6 &  99.6  \\
53551.905 & 1.09 & 12.86-25.15 & 32.02 & 11.08 $\pm$ 0.32 & 3.48 $\pm$ 0.10 & 17.37 $\pm$ 0.51 & 2.09 $\pm$ 0.03 &  26.2/11 & 0.6  \\
53554.906 & 0.68 & 15.21-22.32 & 12.52 & 3.52 $\pm$ 0.30 & 1.11 $\pm$ 0.09 & 5.91 $\pm$ 0.47 & 2.26 $\pm$ 0.11 & 3.9/9  & 92.1  \\
53555.914 & 0.44 & 12.85-22.32 & 6.08 & 1.27 $\pm$ 0.34 & 0.40 $\pm$ 0.11 & 2.96 $\pm$ 0.62 & 1.97 $\pm$ 0.29 & 1.9/6 &  92.5  \\
53557.916 & 0.54 & 12.84-19.06 & 8.40 & 2.25 $\pm$ 0.32 & 0.71 $\pm$ 0.10 & 3.91 $\pm$ 0.48 & 2.30 $\pm$ 0.21 & 6.5/7 &  48.5  \\
53559.920 & 0.98 & 9.94-17.22 & 10.05 & 1.85 $\pm$ 0.23 & 0.58 $\pm$ 0.07 & 3.10 $\pm$ 0.33 & 2.25 $\pm$ 0.13 & 8.4/8 &  39.9  \\
53560.906 & 0.76 & 9.96-19.07 & 24.39 & 9.93 $\pm$ 0.38 & 3.12 $\pm$ 0.12 & 14.35 $\pm$ 0.56 & 2.20 $\pm$ 0.04 &  22.5/11 & 2.1  \\
53562.911 & 1.63 & 9.94-16.79 & 11.08 & 2.19 $\pm$ 0.37 & 0.69 $\pm$ 0.12 & 2.83 $\pm$ 0.30 & 2.34 $\pm$ 0.13 & 14.1/8  & 8.2  \\
53563.921 & 0.85 & 9.94-15.16 & 18.69 & 5.53 $\pm$ 0.28 & 1.74 $\pm$ 0.09 & 7.89 $\pm$ 0.39 & 2.25 $\pm$ 0.06 & 11.5/9  & 24.3  \\
53564.917 & 0.34 & 9.94-15.18 & 8.91 & 2.89 $\pm$ 0.46 & 0.91 $\pm$ 0.15 & 4.88 $\pm$ 0.56 & 2.27 $\pm$ 0.20 & 5.4/6 &  49.7  \\
53565.920 & 2.57 & 9.95-28.93 & 11.62 & 1.71 $\pm$ 0.13 & 0.54 $\pm$ 0.04 & 2.73 $\pm$ 0.22 & 2.49 $\pm$ 0.12 & 10.7/8  & 21.6  \\
53566.953 & 1.91 & 9.99-30.10 & 11.63 & 1.33 $\pm$ 0.11 & 0.42 $\pm$ 0.04 & 2.16 $\pm$ 0.20 & 2.28 $\pm$ 0.13 & 7.4/10  & 69.0  \\

\hline
\enddata

\tablenotetext{a}{Net observation time after removing bad-quality runs.}
\tablenotetext{b}{Zenith Angle range covered during the observation.}
\tablenotetext{c}{Combined significance of detected signal in the 0.1-10 \TeV\ band.}
\tablenotetext{d}{Integrated flux above 0.15 \TeV.}
\tablenotetext{e}{Normalization factor of the PL fit.}
\tablenotetext{f}{Slope of PL fit.}
\tablenotetext{g}{$\chi^2$ value and number of degrees of freedom of the power-law fit.}
\tablenotetext{h}{Chance probability for larger $\chi^2$ values.}

\end{deluxetable}

\begin{deluxetable}{ccccccccccc}
\tabletypesize{\scriptsize}
\tablecaption{Flare model parameters: integral emission above 0.15 \TeV.
\label{fitflareresults}}
\tablehead{
\colhead{Date} &\colhead{$T_{obs}$ \tablenotemark{a}}
&\colhead{$S_{comb}$\tablenotemark{b}} 
& \colhead{$a$\tablenotemark{c}}
& \colhead{$a$}
& \colhead{$b$}
& \colhead{$c$}
& \colhead{$d$}
& \colhead{${\chi^2}/{NDF}$\tablenotemark{d}}
 & \colhead{$P$}\tablenotemark{e} \\
\colhead{} & \colhead{($h$)} & \colhead{(sigma)}
& \colhead{($\frac{10^{-10} ~ph}{cm^2 \cdot s}$)} &\colhead{{\it (Crab Units)}}
& \colhead{($\frac{10^{-10} ~ph}{cm^2 \cdot s}$)} 
& \colhead{$(s)$} & \colhead{$(s)$} 
& \colhead{} &\colhead{(\%)}}
\startdata

June 30 &0.63& 24.7& 10.80$\pm$0.48& 3.39$\pm$0.15&13.2$\pm$4.7   &81$\pm$41&50$\pm$23  &20.0/15&17.3\tablenotemark{f} \\
July 9 & 0.36 & 19.6 & 7.39$\pm$0.48 & 2.32$\pm$0.15& 20.3$\pm$3.3 &95$\pm$24 &185$\pm$40 & 4.2/7 & 75.8 \\
\enddata

\tablenotetext{a}{Net observation time during variable emission (right part of the graphs).}
\tablenotetext{b}{Combined signal significance from variable 
                  emission (right part of the graphs) in 0.1-10 \TeV\ band.}
\tablenotetext{c}{Integrated flux above 0.15 \TeV\ for the steady emission (left part of the graphs).}
\tablenotetext{d}{$\chi^2$ value and number of degrees of freedom of the fit with eq. \ref{eq_fitflare}.}
\tablenotetext{e}{Chance probability of having larger $\chi^2$ values.}
\tablenotetext{f}{If the points after 22:44 are not taken into account, the coefficients from 
the fit remain the same, and the probability increases up to 52.7\% ($\chi^2/NDF=11.0/12$).}
\end{deluxetable}

\begin{deluxetable}{cccccccc}
\tabletypesize{\scriptsize}
\tablecaption{Flare model parameters for June 30: differential emission.
\label{fitflareresults_F0701_ERanges}}
\tablehead{
\colhead{Energy Range}
& \colhead{$a$\tablenotemark{c}}
& \colhead{$a$}
& \colhead{$b$}
& \colhead{$c$}
& \colhead{$d$}
& \colhead{${\chi^2}/{NDF}$\tablenotemark{d}}
 & \colhead{$P$}\tablenotemark{e} \\
\colhead{($\TeV$)}
& \colhead{($\frac{10^{-10} ~ph}{cm^2 \cdot s}$)} &\colhead{{\it (Crab Units)}}
& \colhead{($\frac{10^{-10} ~ph}{cm^2 \cdot s}$)} 
& \colhead{$(s)$} & \colhead{$(s)$} 
& \colhead{} &\colhead{(\%)}}
\startdata

0.25-0.6& 3.30$\pm$0.23& 3.0$\pm$0.2  &7.5$\pm$2.8   &110$\pm$57&61$\pm$26 &5.2/6&51.8 \\

\enddata

\tablenotetext{c}{Integrated steady emission flux (left part of the graphs) in specified energy range.}
\tablenotetext{d}{$\chi^2$ value and number of degrees of freedom of the fit with eq. \ref{eq_fitflare}.}
\tablenotetext{e}{Chance probability of having larger $\chi^2$ values.}
\end{deluxetable}

\begin{deluxetable}{cccccc}
\tabletypesize{\scriptsize}
\tablecaption{Flare model parameters for July 9 resulting from a combined fit to all LCs 
from Fig.  \ref{LCSingle_MultiEnergyRanges_F0709} using 
equation \ref{eq_fitflare} with c=d. The overall ${\chi^2}/{NDF}$ = 14.0/12 ($P$=0.3)
\label{fitflareresults_F0709_ERanges}}
\tablehead{
\colhead{Energy Range}
& \colhead{$a$\tablenotemark{a}}
& \colhead{$a$}
& \colhead{$b$}
& \colhead{$c$}
& \colhead{$t_{0}-t_{0}^{LC~E~0.15-0.25 TeV}$\tablenotemark{b}} \\
\colhead{($\TeV$)}
& \colhead{($\frac{10^{-10} ~ph}{cm^2 \cdot s}$)} &\colhead{{\it (Crab Units)}}
& \colhead{($\frac{10^{-10} ~ph}{cm^2 \cdot s}$)} 
& \colhead{$(s)$}
& \colhead{$(s)$}}
\startdata

0.15-0.25& 4.23$\pm$0.49&2.48$\pm$0.28   &8.6$\pm$3.7  &143$\pm$92 & 0 $\pm$ 68\\
0.25-0.6& 2.55$\pm$0.24&2.32$\pm$0.09   &9.3$\pm$2.5  &95$\pm$28  & 7 $\pm$ 36\\
0.6-1.2 & 0.53$\pm$0.10&1.96$\pm$0.37   &2.7$\pm$0.9  &146$\pm$56  & 111 $\pm$ 91\\
1.2-10  & 0.23$\pm$0.06&1.51 $\pm$0.39 &4.0$\pm$0.9  &103$\pm$19  & 239 $\pm$ 40\\

\enddata
\tablenotetext{a}{Integrated steady emission flux (left part of the graphs) in specified energy range.}

\tablenotetext{b}{$t_{0}^{LC~E~0.15-0.25 TeV}$ is the $t_0$ for the LC in the energy range 0.15-0.25 TeV. 
This is used as a reference value, and the error of this quantity is not taken into account.}

\end{deluxetable}

\begin{deluxetable}{cccccc}
\tabletypesize{\scriptsize}
\tablecaption{Flare model parameters for July 9 resulting from a combined fit to all LCs 
from Fig.  \ref{LCSingle_MultiEnergyRanges_F0709_commont0} using 
equation \ref{eq_fitflare} with $c=d$, and with a common $t_0$ for 
all LCs. The overall ${\chi^2}/{NDF}$ = 25.6/15 ($P$=0.04)
\label{fitflareresults_F0709_ERanges_commont0}}
\tablehead{
\colhead{Energy Range}
& \colhead{$a$\tablenotemark{a}}
& \colhead{$a$}
& \colhead{$b$}
& \colhead{$c$}
& \colhead{$t_{0}-t_{0}^{LC~E~0.15-0.25 TeV}$\tablenotemark{b}}\\
\colhead{($\TeV$)}
& \colhead{($\frac{10^{-10} ~ph}{cm^2 \cdot s}$)} &\colhead{{\it (Crab Units)}}
& \colhead{($\frac{10^{-10} ~ph}{cm^2 \cdot s}$)} 
& \colhead{$(s)$}
& \colhead{$(s)$}} 
\startdata

0.15-0.25 & 4.23$\pm$0.49&2.48$\pm$0.28   &5.4$\pm$2.2  &301$\pm$210 & 0 $\pm$ 42\\
0.25-0.6 & 2.55$\pm$0.24&2.32$\pm$0.09   &5.7$\pm$1.5  &162$\pm$63  & 0 $\pm$ 42\\
0.6-1.2  & 0.53$\pm$0.10&1.96$\pm$0.37   &2.6$\pm$0.8  &153$\pm$56 & 0 $\pm$ 42\\
1.2-10   & 0.23$\pm$0.06&1.51 $\pm$0.39 &3.9$\pm$1.0  &97$\pm$22 & 0 $\pm$ 42\\

\enddata

\tablenotetext{a}{Integrated steady emission flux in specified energy range (left part of the graphs).}

\tablenotetext{b}{$t_{0}^{LC~E~0.15-0.25 TeV}$ is the $t_0$ for the LC in the energy range 0.15-0.25 \TeV. 
This is used as a reference value, and the error of this quantity is not taken into account.}

\end{deluxetable}

\begin{deluxetable}{cccccccccc}
\tabletypesize{\scriptsize}
\tablecaption{Stacked analysis: mean spectral parameters.
\label{FluxlevelsTable}}
\tablehead{
\colhead{Flux Level\tablenotemark{m}} &\colhead{$T_{obs}$ \tablenotemark{a}}&\colhead{ZA\tablenotemark{b}}
&\colhead{$S_{comb}$\tablenotemark{c}} 
& \colhead{$F_{>0.15~TeV}$\tablenotemark{d}}
& \colhead{$F_{>0.15~TeV}$}
& \colhead{$K_0$\tablenotemark{e}}
& \colhead{$a$\tablenotemark{f}} 
& \colhead{${\chi^2}/{NDF}$\tablenotemark{g}} & \colhead{$P$\tablenotemark{h}} \\
\colhead{} & \colhead{($h$)} & \colhead{($deg$)} & \colhead{sigma}
& \colhead{($\frac{10^{-10} ~ph}{cm^2 \cdot s}$)} &\colhead{{\it (Crab Units)}}
& \colhead{($\frac{10^{-10} ~ ph}{cm^2 \cdot s \cdot 0.3 TeV}$)}
& \colhead{} & \colhead{} &\colhead{(\%)}}
\startdata
$Low$ & 17.2&  9.96-30.1&16.7&	1.24$\pm$ 0.08& 0.39 $\pm$ 0.02&2.31$\pm$0.13&2.45 $\pm$ 0.07 & 7.8/7&34.6 \\
$Medium$ &11.0&9.95-30.0&22.8&	2.11$\pm$ 0.09&	0.66 $\pm$ 0.03 &3.57$\pm$0.15& 2.43 $\pm$ 0.05 &2.9/7&  89.4 \\
$High$   &1.52&9.95-22.3&21.7&	4.62$\pm$ 0.21&	1.45 $\pm$ 0.07&7.13$\pm$0.32 & 2.28 $\pm$ 0.05 &4.8/7&  68.7 \\

\hline
\enddata
\tablenotetext{m}{See section \ref{Sed4} for definition of flux levels.}
\tablenotetext{a}{Net observation time after removing bad quality runs.}
\tablenotetext{b}{Zenith Angle range covered during the observation.}
\tablenotetext{c}{Combined significance of detected signal in energy range 0.10-TeV-10 \TeV.}
\tablenotetext{d}{Integrated flux above 0.15 \TeV.}
\tablenotetext{e}{Normalization factor of the power-law fit.}
\tablenotetext{f}{Slope of the power-law fit.}
\tablenotetext{g}{$\chi^2$ value and number of degrees of freedom of the power-law fit.}
\tablenotetext{h}{Chance probability of having $\chi^2$ values.}

\end{deluxetable}

\begin{deluxetable}{cccccccccc}
\tabletypesize{\scriptsize}
\tablecaption{PL (eq. \ref{eq_powerlaw}) and Log-parabolic (eq. \ref{eq_powerlawEDepInd}) 
fit results for June 30 and July 9: night-integrated spectra.
\label{FlareFitComparisonTable}}
\tablehead{
\colhead{} 
&\multicolumn{4}{c}{Fit performed with eq. \ref{eq_powerlaw}}
& \multicolumn{5}{c}{Fit performed with eq. \ref{eq_powerlawEDepInd}} \\
\hline 
\colhead{Date} 
& \colhead{$K_0$}
& \colhead{$a$}
& \colhead{${\chi^2}/{NDF}$}
& \colhead{$P$}\tablenotemark{h} 
& \colhead{$K_0$}
& \colhead{$a$} 
& \colhead{$b$} 
& \colhead{${\chi^2}/{NDF}$} 
& \colhead{$P$}\tablenotemark{h} \\
\colhead{} 
& \colhead{($\frac{10^{-10} ~ ph}{cm^2 \cdot s \cdot 0.3 TeV}$)}
& \colhead{} & \colhead{} &\colhead{(\%)}
& \colhead{($\frac{10^{-10} ~ ph}{cm^2 \cdot s \cdot 0.3 TeV}$)}
& \colhead{} & \colhead{} & \colhead{} &\colhead{(\%)}}

\startdata

June 30 & 17.4$\pm$0.05& 2.09$\pm$0.03 &26.1/11& 0.6 & 18.6$\pm$0.06& 1.89$\pm$0.06& 0.35$\pm$0.09 &6.1/10& 80.1 \\ 
July 9 & 14.3$\pm$0.06& 2.20$\pm$0.04 &22.5/11& 2.1 & 15.5$\pm$0.07& 2.06$\pm$0.07& 0.36$\pm$0.16 &15.2/10& 12.5 \\ 

\hline
\enddata

\tablenotetext{h}{Chance probability for larger $\chi^2$ values.}

\end{deluxetable}

\begin{deluxetable}{cccccccccc}
\tabletypesize{\scriptsize}
\tablecaption{Log-parabolic (eq. \ref{eq_powerlawEDepInd}) fit results for June 30 and July 9, 2005: pre-burst and burst spectra.
\label{FlareNightsSplitSEDTable}}
\tablehead{
\colhead{MJD} &\colhead{$T_{obs}$ \tablenotemark{a}}
&\colhead{$S_{comb}$\tablenotemark{c}} 
& \colhead{$F_{>0.15~TeV}$\tablenotemark{d}}
& \colhead{$F_{>0.15~TeV}$}
& \colhead{$K_0$}
& \colhead{$a$}
& \colhead{$b$} 
& \colhead{${\chi^2}/{NDF}$}
& \colhead{$P$}\tablenotemark{h} \\
\colhead{Start} & \colhead{($h$)} 
& \colhead{sigma}
& \colhead{($\frac{10^{-10} ~ph}{cm^2 \cdot s}$)} &\colhead{{\it (Crab Units)}}
& \colhead{($\frac{10^{-10} ~ ph}{cm^2 \cdot s \cdot 0.3 TeV}$)}
& \colhead{} & \colhead{} & \colhead{} &\colhead{(\%)}}

\startdata

53551.905 & 0.46 & 22.3 &10.99$\pm$0.48&3.46$\pm$0.15& 19.8$\pm$1.0& 1.97$\pm$0.08 & 0.27$\pm$0.14 &8.2/9& 51.2 \\ 
53551.924 & 0.63 & 24.7 &11.15$\pm$0.43&3.50$\pm$0.14& 17.2$\pm$0.8& 1.87$\pm$0.08 & 0.34$\pm$0.13 &13.8/10& 18.1  \\  
\hline
53560.906 & 0.40 & 15.2 &  7.64 $\pm$ 0.48 & 2.40 $\pm$ 0.15 &12.7$\pm$1.1& 2.11$\pm$0.12 & 0.57$\pm$0.34 &6.4/8&  59.8 \\  
53560.923 & 0.36 & 19.6 &  12.39 $\pm$ 0.60 & 3.89 $\pm$ 0.19 &19.3$\pm$1.3& 2.00$\pm$0.10 & 0.44$\pm$0.23 &8.9/8&  35.2 \\  

\hline
\enddata

\tablenotetext{a}{Net observation time after removing bad-quality runs.}
\tablenotetext{c}{Combined significance of detected signal in energy \mbox{range 0.1-10 \TeV}.}
\tablenotetext{d}{Integrated \gray\ flux above 0.15 \TeV.}
\tablenotetext{h}{Chance probability for larger $\chi^2$ values.}

\end{deluxetable}

\begin{deluxetable}{cccccccccc}
\tabletypesize{\scriptsize}
\tablecaption{SSC model parameters
\label{SSCModelParams}}
\tablehead{
\colhead{spectrum} 
& \colhead{$\gamma_{\rm min}$}
& \colhead{$\gamma_{\rm br}$}
& \colhead{$\gamma_{\rm max}$}
& \colhead{n1}
& \colhead{n2}
& \colhead{B}
& \colhead{K}
& \colhead{R}
& \colhead{Doppler factor} \\
\colhead{} 
& \colhead{}
& \colhead{}
& \colhead{}
& \colhead{}
& \colhead{}
& \colhead{Gauss} 
& \colhead{particle/$cm^3$} 
& \colhead{cm} 
& \colhead{}} 
\startdata
June 30         & 1 & $10^6$         & $10^7$         & 2 & 3.5 & 0.23& 7.5$\cdot$ $10^4$ & $10^{15}$ & 25 \\ 
June 30 (bis)   & 1 & $5 \cdot 10^5$ & $10^7$         & 2 & 3.5 & 0.053 & 7.0$\cdot$ $10^4$ & $10^{15}$ & 50 \\
{\it Low flux}  & 1 & $10^5$         & $5 \cdot 10^6$ & 2 & 3.2 & 0.31 & 4.3$\cdot$ $10^4$ & $10^{15}$ & 25 \\

\hline
\enddata

\end{deluxetable}

\cleardoublepage

\cleardoublepage

\clearpage

\end{document}